\documentclass[graybox,natbib,nosecnum]{svmult}
\bibpunct{(}{)}{;}{a}{}{,} 

\pdfoutput=1   

\usepackage{mathptmx}       
\usepackage{helvet}         
\usepackage{courier}        
\usepackage{type1cm}        

\usepackage{makeidx}         
\usepackage{graphicx}        
\usepackage{multicol}        
\usepackage[bottom]{footmisc}
\usepackage[normalem]{ulem}	
\usepackage{hyperref}  

\usepackage{soul}   

\definecolor{darkgreen}{RGB}{0,142,128}

\newcommand*\aap{A\&A}

\newcommand*\apj{ApJ}
\newcommand*\apjl{ApJ}

\newcommand*\apjs{ApJS}

\newcommand*\apss{Ap\&SS}

\newcommand*\jgr{J Geophys Res}

\newcommand*\mnras{MNRAS}

\newcommand*\nat{Nature}

\newcommand*\planss{Planet Space Sci}

\newcommand*\ssr{Space Sci Rev}

\newcommand{\hbindex}[1]{#1}  

\makeindex             

\begin{document}

\title*{Models of Star-Planet Magnetic Interaction}
\author{Antoine Strugarek}
\institute{Laboratoire AIM,
  DRF/IRFU/SAp, CEA Saclay, F-91191 Gif-sur-Yvette Cedex, France,
  \email{antoine.strugarek@cea.fr} \and D\'epartement de Physique, Universit\'e de Montr\'eal,
  C.P. 6128 Succ. Centre-ville, Montr\'eal, Qu\'ebec, H3C 3J7, Canada,
  \email{strugarek@astro.umontreal.ca}}

%
%
\maketitle

\abstract{Magnetic interactions between a planet and its environment
  are known to lead to phenomena such as aurorae and shocks in the solar system. The
  large number of \hbindex{close-in exoplanets} that were discovered triggered a
  renewed interest in magnetic interactions in star-planet
  systems. Multiple other magnetic effects were then unveiled, such as planet
  inflation or heating, planet migration, planetary material escape, and
  even modification of the host star properties. We review here the
  recent efforts in modelling and understanding magnetic interactions
  between stars and planets in the context of compact
  systems. We first provide simple estimates of the
  effects of magnetic interactions and then detail analytical and
  numerical models for different representative scenarii. We finally
  lay out a series of future developments that are needed today to better
  understand and constrain these fascinating interactions.
}

\section{Introduction }

Stars and planet interact mainly through gravitation, magnetic fields
and radiation. In this review we will focus on \hbindex{star-planet magnetic
interaction} (\hbindex{SPMI}) for close-in planets around cool stars. By close-in planet
we mean here planets that are sufficiently close to their star to
orbit in a region where the \hbindex{wind} of the star is in a
sub-alfv\'enic regime (\textit{i.e.} the local speed of the \hbindex{wind} is smaller than the
local Alfv\'en speed). Tidal and radiative interactions will be covered
in other chapters of this book.

Numerous intriguing observations related to close-in systems have been
reported with the advent of modern space telescopes and ground-based instruments
(SPITZER, CoRoT, Kepler/K2, HARPS, HIRES, SPHERE, ...). To mention a few, these
observations report anomalous chromospheric activity in close-in
planet hosting stars \citep{Shkolnik:2008gw,Poppenhaeger:2014be}, a lack of X-ray emission
in WASP-12 \citep{Fossati:2013io} and WASP-18
\citep{Pillitteri:2014jy}, possible bow shock absorption in
the UV for HD 189733 \citep{Llama:2013il,Cauley:2015kl,Turner:2016bp}, a dearth of
close-in planets around fast rotating stars
\citep{Pont:2009ip,McQuillan:2013jw,Lanza:2014hw}, and much
more \citep[\textit{see,
  \textit{e.g.}}][]{Miller:2015ih,2016A&A...592A.143F,Staab:2017kq,Mengel:2016kk}. Magnetic
interactions are today a serious candidate to explain these
fascinating phenomena.

We concentrate here on the effects of magnetic interactions and present
the recent theoretical efforts for modelling them. We first review the
main awaited effects of \hbindex{SPMI}s and give an order of magnitude estimate for each of
them. Then, we distinguish the cases of un-magnetized and magnetized
planets and successively review analytical and numerical modelling
efforts for each case. We conclude by listing the model improvements 
that are needed today for our understanding of \hbindex{SPMI}s, and for helping
the interpretation of future exoplanetary systems observations.

\section{General characteristics of star-planet magnetic interactions}
\label{sec:gener-char-star}

\begin{figure}[!htbp]
  \centering
  \includegraphics[width=0.8\linewidth]{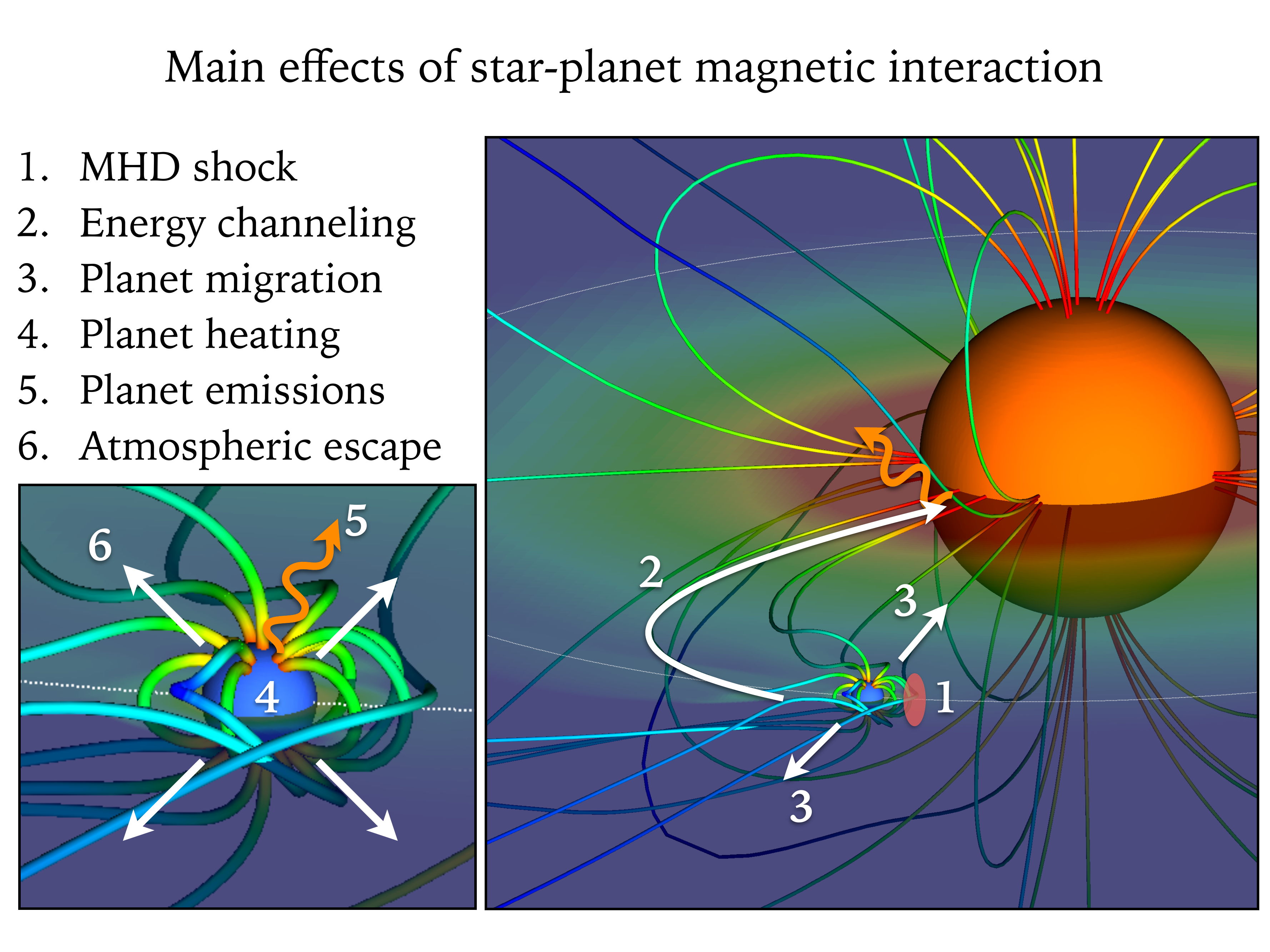}
  \caption{Summary of the main effects of star-planet magnetic
    interactions discussed in this review. The illustrating image is
    based on a model published in \citet{Strugarek:2015cm} and shows a
  close-in planet (small blue sphere) around its host (orange sphere).
The magnetic field lines are colored according to the magnetic field strength, and
the transparent coloured plane shows the plasma density.}
  \label{fig:effects_SPMI}
\end{figure}

In this section we will describe the main possible impacts of
star-planet magnetic interactions in distant exoplanetary systems,
which are summarized in Figure \ref{fig:effects_SPMI}. 
Detailed models
of \hbindex{SPMI} will be discussed in subsequent sections.

\subsection{Main effects of star-planet magnetic interactions}
\label{sec:main-outcome-star}

Solar-type stars are thought
to generate their large-scale magnetic field \citep{Donati:2009if}
through dynamo processes in their 
convection zone \citep[][]{Brun:2015kc,Brun:2015co}. This
magnetic field shapes 
the environment in which close-in planets orbit. Magnetic
interactions then develop due to the orbital motion,
as long as
the planet is composed --at least in part-- of ionized
material. Multiple effects can occur due to the magnetic interaction, which
we list hereafter. It is nevertheless important to realize that
\hbindex{SPMI}s are generally time-dependent
and susceptible to intermittency, since close-in planets
encounter inhomogeneous environments along their orbit.

\runinhead{1. Magneto-hydrodynamic \hbindex{shock}} 
The relative motion ${\bf v}_o = {\bf v}_{K} -
{\bf v}_w$ between the ambient
plasma (${\bf v}_w$) and the orbiting planet (keplerian velocity ${\bf
  v}_K$) can be super-Alfv\'enic due to the proximity of the planet
to its host. 
Under the assumption of a circularized orbit, the
Keplerian velocity of the planet can be approximated by $v_{K} \simeq
\sqrt{GM_\star/R_{\rm orb}}$ (where $R_{\rm orb}$ is the orbital radius). Close to the
star the \hbindex{wind} speed in the orbital direction is likely to be
rotationally constrained by the rotating host, and may be written $v_w
\simeq R_{\rm orb} \Omega_\star$ ($\Omega_\star$ is the stellar rotation
rate).
The coronal density can be
supposed to decrease with orbital distance as a power-law, \textit{i.e.} $\rho \simeq \rho_\star
(R/R_\star)^{-\alpha}$. The decrease of the coronal density is not
well constrained by models or observations today, here to simplify the discussion we will
crudely assume
$\alpha=8$, which is a fair approximation to density profiles obtained
in standard 3D \hbindex{stellar wind} models very close to the star. Assuming a dipolar field for the stellar magnetic field (for the sake of simplicity, defined by a stellar Alfv\'en speed
$v_{A\star}$ at the base of its corona),
the \hbindex{shock} condition $v_o > v_A$ can be written as
\begin{equation}
  \label{eq:shock_crit_2}
  R_{\rm orb} < R_{\rm crit} = R_\star \left(f+\upsilon \right)^{-1/2}\, ,
\end{equation}
where $R_{\rm orb}$ is the orbital radius,
$f=R_\star\Omega_\star\left(GM_\star/R_\star\right)^{-1/2}$ is the
keplerian-normalized measure of the stellar rotation rate and $\upsilon =
v_{A\star}\left(GM_\star/R_\star\right)^{-1/2}$ a normalized measure
of the stellar magnetic field. 
If the star rotates rapidly (large $f$) or possesses a
strong magnetic field (large $\upsilon$), a close-in planet will
likely not possess a bow-shock.
For the particular case of the Sun, we expect
$f_\odot \simeq 6 \times 10^{-4}$ and $\upsilon_\odot \simeq 0.1 - 10$,
which gives a critical radius between $0.3$ and $3$ solar
radii. Hence, only an extremely close-in planet ($R_{\rm orb} < 3
R_\odot$) could in principle develop a bow-shock in the sub-alfv\'enic region of
a solar-like \hbindex{wind}. As the orbital radius increases, the various
approximations used to derive Equation \ref{eq:shock_crit_2} become
invalid and the orbital radius eventually crosses the Alfv\'en surface
of the stellar \hbindex{wind}, where a shock will (almost) systematically develop. 

\begin{figure}[!htbp]
  \centering
  \includegraphics[width=\linewidth]{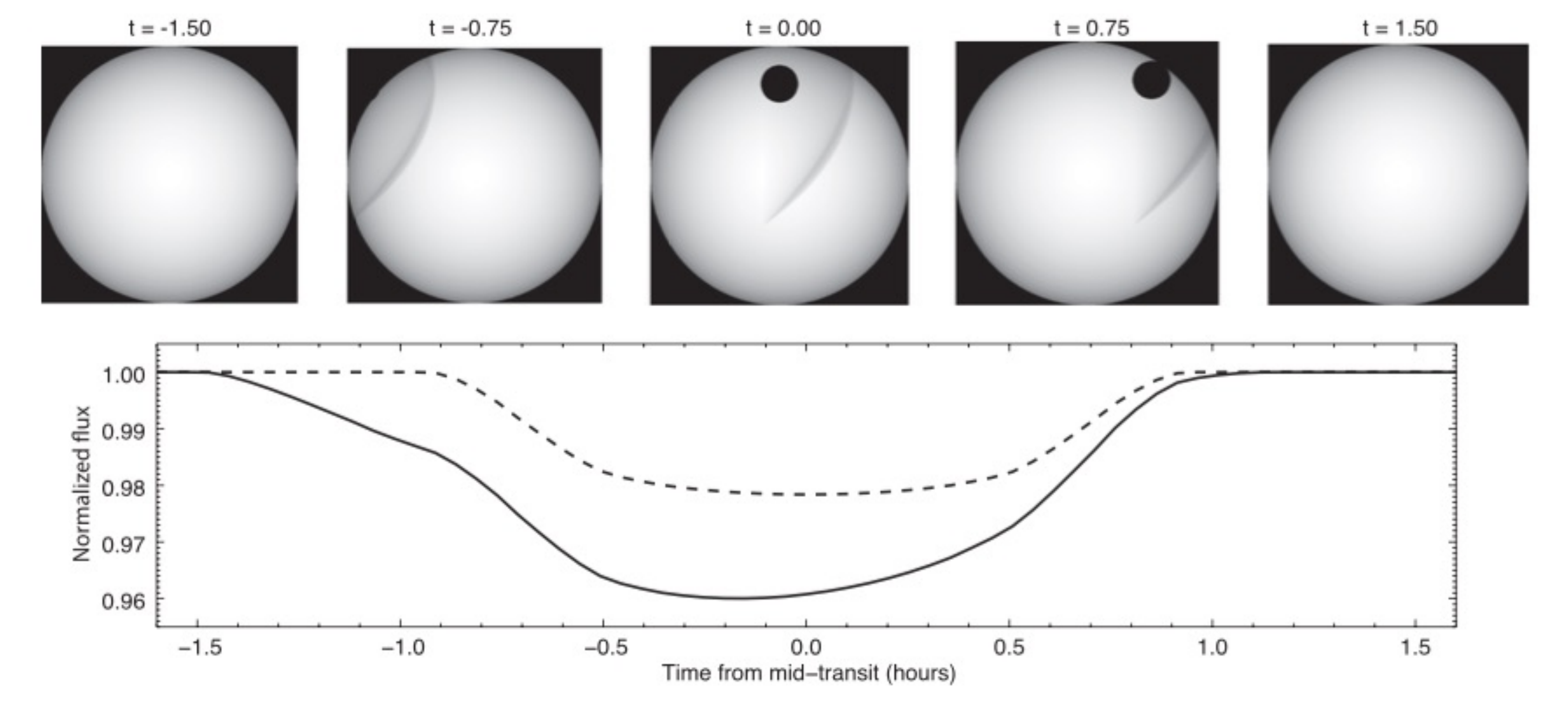}
  \caption{\hbindex{Shock} model during the transit of H189733b. The shock front
  affects the transit (bottom panel, solid line) compared to the case with no
  shock (dashed line). Figure adapted from \citet{Llama:2013il}.}
  \label{fig:shock_example}
\end{figure}

One of the most interesting aspect of the development of a \hbindex{shock} in
close-in systems is
the possibility (at least in theory) to actually observe its trace for
transiting planets. This idea was recently put forward by
\citet{Llama:2013il,Cauley:2015kl} in the context of the HD 189733 system (see
Figure \ref{fig:shock_example}). At the
nose of the shock, material accumulation can cause a
localized high density region. Prior to a transit, such a shock could
in principle lead to an excess absorption of the stellar luminosity
in several wavelengths (typically in the visible and near-UV
spectra). 
They find that the shock position
deduced from the pre-transit absorption suggests a planetary magnetic
field strength of about 28 G (approximately 7 times larger than
Jupiter's magnetic field). Even though many assumptions were made to
deduce this value \citep[see also][]{Turner:2016kp}, pre-transit absorption observations
remain a promising technique 
for the difficult task of constraining exoplanetary magnetic fields.

\runinhead{2. Magnetic energy channeling} A close-in planet can be viewed
as a perturber orbiting in the likely non-axisymmetric inter-planetary
medium. The perturbations will take the form of magneto-sonic waves,
travelling away from the planet location in the (${\bf v}_o$, ${\bf B}_w$)
plane (where ${\bf B}_w$ is the interplanetary magnetic field). 
The degeneracy of the group velocity
of the standard Alfv\'en waves (the group velocity is independant of
the perturbation direction) allows for the focused propagation of waves
packets along the Alfv\'en characteristics. This results in a net
\hbindex{Poynting flux} channeled away from the planet, along what is often
referred to as \textit{\hbindex{Alfv\'en wings}}. One may do a back of the envelope calculation
to estimate the travel time of Alfv\'en
waves between the planet and the star. Alfv\'enic perturbations can
travel back and forth the star and the planet if 
\begin{equation}
  \label{eq:travel_back_forth}
  \left|\frac{f-\left(R_{\rm
          orb}/R_\star\right)^{-3/2}}{\left(R_{||}/R_\star\right)\left(R_{\rm
            orb}/R_\star\right)^{-5}\left(\upsilon/2I_E\right)} \right| < 1 \, ,
\end{equation}
where $R_{||}$ is the obstacle size along ${\bf v}_0$, $I_E$ is a geometric
integral along the magnetic field lines, and we have assumed a classical alfv\'enic
profile of a Weber and Davis-like solar \hbindex{wind} (see previous
paragraph). Using a solar-twin as an example, it appears that unless the
magnetosphere of the planet is very large (typically of the order of
the Sun itself), the perturbations
triggered by the planet do not have time to travel back and forth
between the planet and the star. This situation corresponds to the
so-called \textit{pure \hbindex{Alfv\'en wing}} case
\citep{Neubauer:1998bw}. Other scenarii can be
realized, depending on the propagation time of these waves, the
interested reader may find a detailed analytical description of them
in \citet{Saur:2017wi}. Nevertheless, in all cases the energy flux
carried by the waves propagates
in the (${\bf v}_o$, ${\bf B}_w$) plane in the form of two wings. Depending
on ${\bf v}_o$ and ${\bf B}_w$, both wings can connect onto host
star; only one may while the other extends away from the star towards the interplanetary
medium; or both may head away from the host
star. As a result, the knowledge of the magnetic configuration in
between the host star and the orbital path of the planet is mandatory
to assess how much energy can actually be channeled (and where
exactly) onto the host.

\begin{figure}[!htbp]
  \centering
  \includegraphics[width=0.5\linewidth]{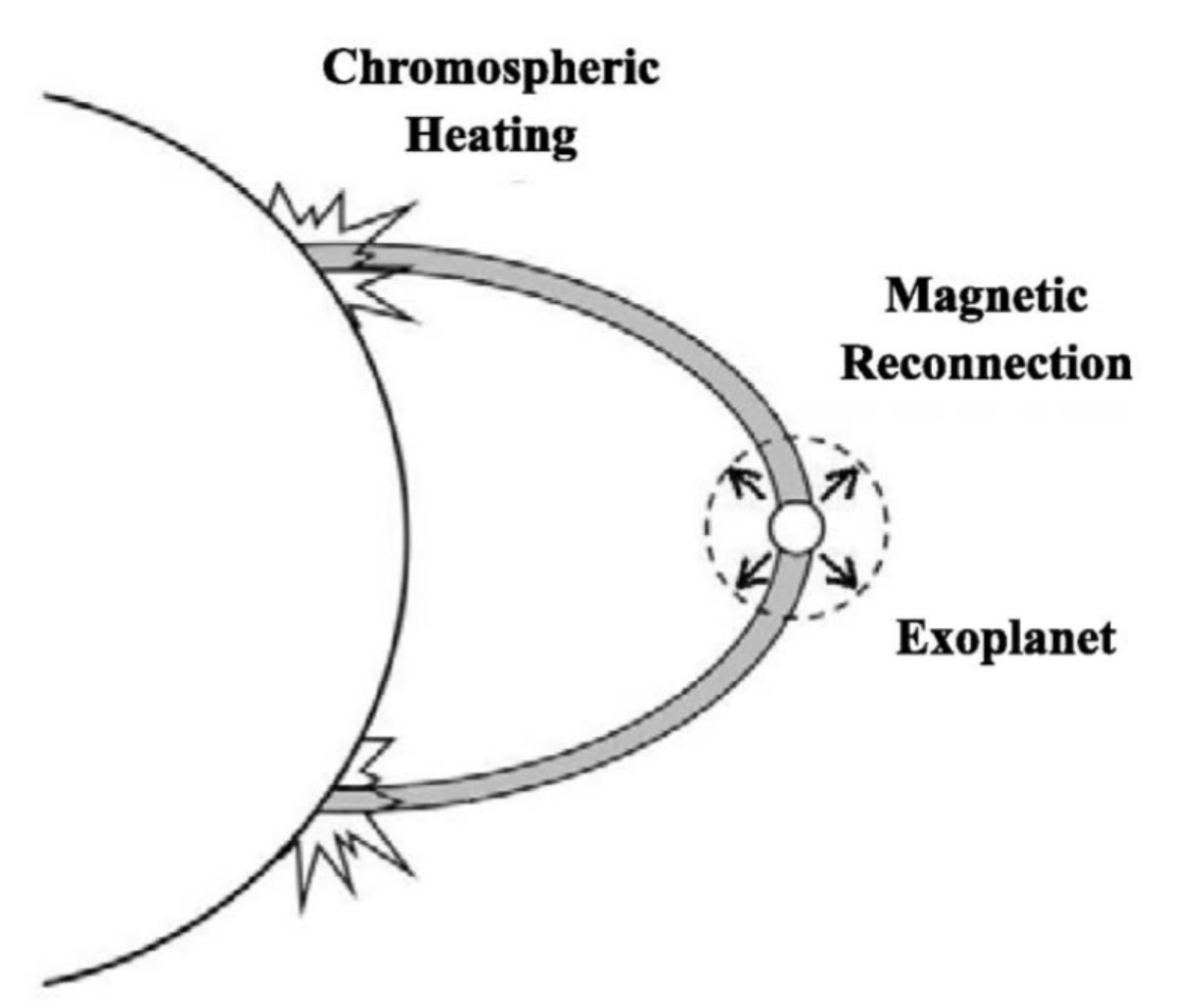}
  \caption{Cartoon of the energy channeling due to the \hbindex{SPMI}, from the
    pioneering modelling work of \citet{Ip:2004ba}. Reproduced with
    the permission of AAS.}
  \label{fig:e_channeling_example}
\end{figure}

The idea of observable traces of this energy flux in exoplanetary
systems traces back to early 2000's
\citep[\textit{e.g.}][]{Cuntz:2000ef,Rubenstein:2000hp} through the
form of stellar activity enhancement at the impact point of the energy
flux (see Figure \ref{fig:e_channeling_example}). Since then a handful
of detections of anomalous activity correlated with
the planet orbital period were reported \citep[see
\textit{e.g.}][]{Shkolnik:2008gw}. It is nonetheless important to
realize that the impact point of the Poynting flux on the stellar
chromosphere is determined by both ${\bf v}_o$ \textbf{and} ${\bf B}_w$. If
the stellar magnetic field is an inclined dipole, for instance, the
impact point will at first order circulate around the magnetic pole of the star as
the planet orbits around its host, and the enhanced emissions
associated with the \hbindex{SPMI} will be correlated with the stellar rotation
rather than the orbital period. Conversely, if the stellar magnetic
field is a dipole perfectly perpendicular to the orbital plane, the
enhanced emissions will be correlated with the orbital period. 

\runinhead{3. \hbindex{Planet migration} and host star spin up/down} 
The planet (with its
magnetosphere, if any) can be viewed as an obstacle in a flow and
consequently suffers a drag force from the ambient medium. The angular momentum lost by the
planet will generally be exchanged with its host, spinning up/down the
central star. Because the interplanetary medium is magnetized and the
planet may possess a magnetosphere, the drag force felt by the planet
depends as well on the magnetic topology of the interaction (these
aspects will be detailed in the next sections). 
The direction of the angular
momentum exchange can then be estimated by an order of magnitude
calculation similar to
Equation (\ref{eq:shock_crit_2}). The planet will migrate outwards
only if
\begin{equation}
  \label{eq:dir_amom}
  R_\star f^{-2/3} < R_{\rm orb} < R_A\, ,
\end{equation}
where $R_A$ is the Alfv\'en radius on the orbital plane. Again using
the Sun as an example, we find that the orbital radius has to be at the
same time larger than $140$ solar radii and smaller than $R_A \simeq
15 R_\odot$, which of course cannot be realized. Hence, close-in
planets orbiting solar twins will
systematically lose their orbital angular momentum due to \hbindex{SPMI} and
inexorably fall onto their host if no other physical process sets
in. Only planets around very fast rotators may experience outward
\hbindex{migration} due to \hbindex{SPMI}s.

\begin{figure}[!htbp]
  \centering
  \includegraphics[width=0.8\linewidth]{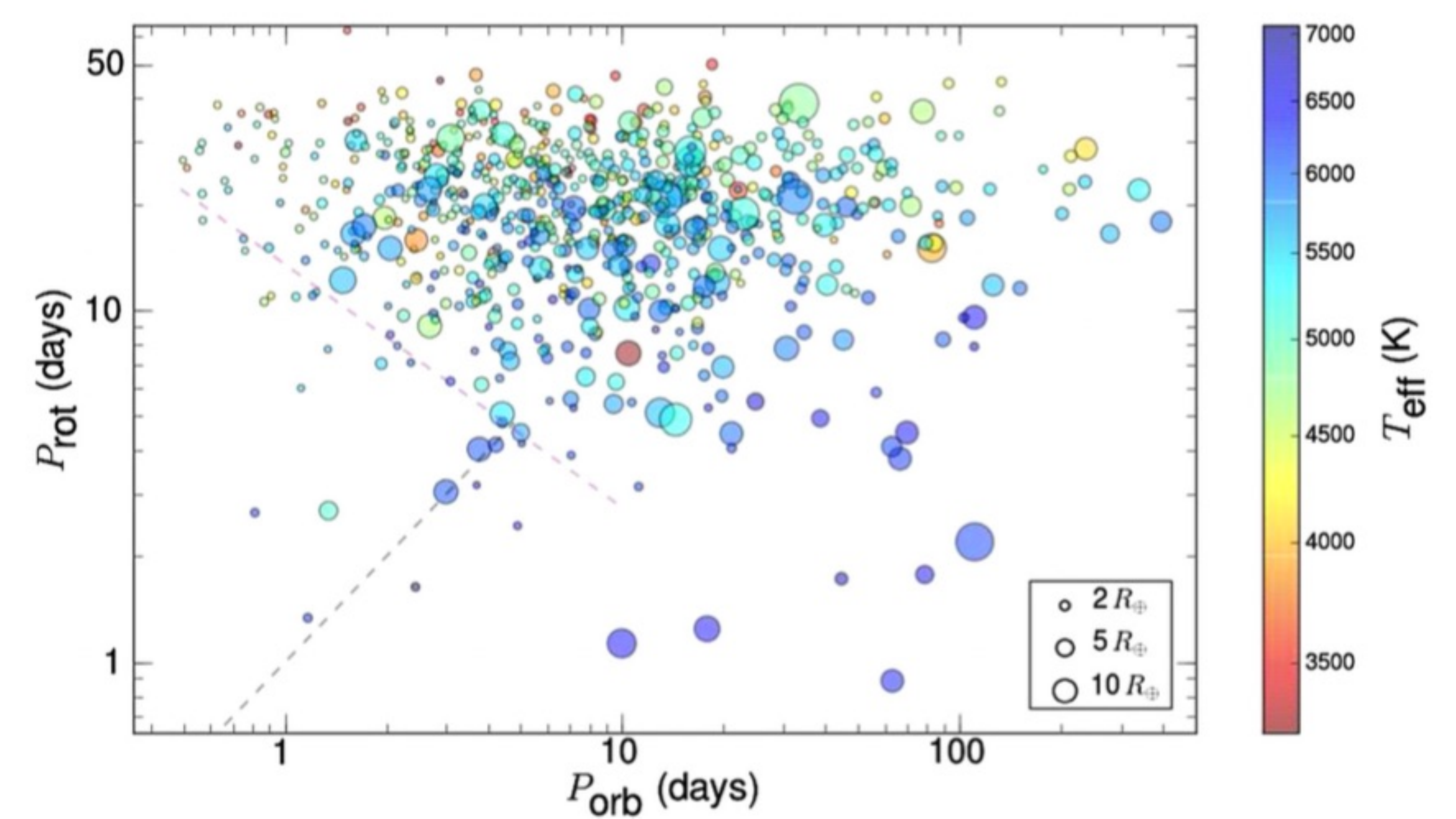}
  \caption{Distribution of known exoplanets as a function of orbital
    period and rotation period of their host. The dearth of exoplanets
    nearby fast rotators clearly appears in the bottom left corner of
    the picture. Adapted from \citet{McQuillan:2013jw}. Reproduced with
    the permission of AAS.}
  \label{fig:distribution_exoplanets}
\end{figure}

The strength of the magnetic torque $\mathcal{T}$ felt by the planet is directly controlled by both the
large-scale magnetic field of the star, and the size of the obstacle
composed of the planet and its magnetosphere. It can be generically
written as
\begin{equation}
  \label{eq:torque}
  \mathcal{T} = c_d R_{\rm orb} A_{\rm eff} P_t\, ,
\end{equation}
where $c_d$ is a drag coefficient which represents the efficiency of
the magnetic coupling, $P_t$ is the total pressure of the ambient
plasma impacting the planetary obstacle (generally, $P_t$ will be
dominated by the magnetic pressure in the stellar \hbindex{wind} for close-in
systems, see \textit{e.g.} \citealt{Strugarek:2016ee}), $A_{\rm eff}$
the effective area of the planetary obstacle, and as a result $R_{\rm
  orb} A_{\rm eff} P_t$ is the amount of angular momentum that can be
transfered to/from the planet orbital motion, with an efficiency
parameter $c_d$. 
For T Tauri stars with typical magnetic
fields of the order of $10^4$ G, the \hbindex{migration} time-scale associated
with the torque $\mathcal{T}$ can be of the order of 100 Myr
\citep{Strugarek:2015cm}. 
As a result, magnetic torques have to be
taken into account to explain the observed population of close-in planets
with respect to the rotation period of their host
\citep[\textit{e.g.}][see Figure \ref{fig:distribution_exoplanets}]{Pont:2009ip,McQuillan:2013jw,Lanza:2014hw,Damiani:2015ef}.

\runinhead{4. Planet heating} In the case where the interplanetary field
is able to permeate into at least a part of the planet, ohmic
dissipation inside the planetary body may lead to a substantial
heating \citep[\textit{e.g.}][]{Laine:2008dx,Laine:2012jt}. Such a
dissipation could in theory lead to planet inflation, possibly result
in a planetary mass loss due to Roche lobe overflow, or even
molecular dissociation. This effect, though, directly depends on the resistivity
profile inside the planet, which is very poorly constrained for
close-in objects as of today. 

\runinhead{5. Extreme events, aurorae and planetary emissions} A
by-product of the magnetic interaction is also the possibility of
emissions inside the planetary magnetosphere, alike aurorae on Earth. These aurorae can be multi-wavelength signals, and are
expected to be particularly intense in the radio domain
\citep[\textit{e.g.}][]{Zarka:2007fo,Griessmeier:2007dm}. They can be
sustained due to the continuous interaction of the planet with the
ambient \hbindex{wind}, or due to particular extreme eruptive events triggered
in the stellar lower corona and impacting the exoplanet. It is fairly unlikely
that we will be able anytime soon to capture the signature of the latter case,
hence researchers have focused on characterizing the continuous radio
emission expected from the \hbindex{SPMI} \citep[see][]{Zarka:2017uu}.

\runinhead{6. Atmospheric escape} The stellar irradiation of the planet
outer layers can lead to a substantial outflow \citep{2013MNRAS.428.2565T,Matsakos:2015ju,Khodachenko:2015kh} and leave observable
signatures for a distant observer. This phenomenon is not directly
related to magnetic interactions, but magnetic fields can mediate and
alter these outflows when the gas composing it is significantly
ionized \citep{2011ApJ...730...27A}. Hence, we simply mention this
effect here and defer the reader to \citet{Barman:2017wa} for in-depth discussion of this phenomenon. 


\subsection{Stellar wind and star-planet magnetic
  interactions}
\label{sec:import-stell-wind}

We saw that most of the effects of the \hbindex{SPMI}s heavily depend on the
plasma conditions in the \hbindex{stellar wind} at the orbital position as well
as on the path in between the star and the planet. As a result,
\hbindex{stellar wind} models can be used on their own to infer valuable
informations about \hbindex{SPMI}s.

\begin{figure}[!htbp]
  \centering
  \includegraphics[width=0.4\linewidth]{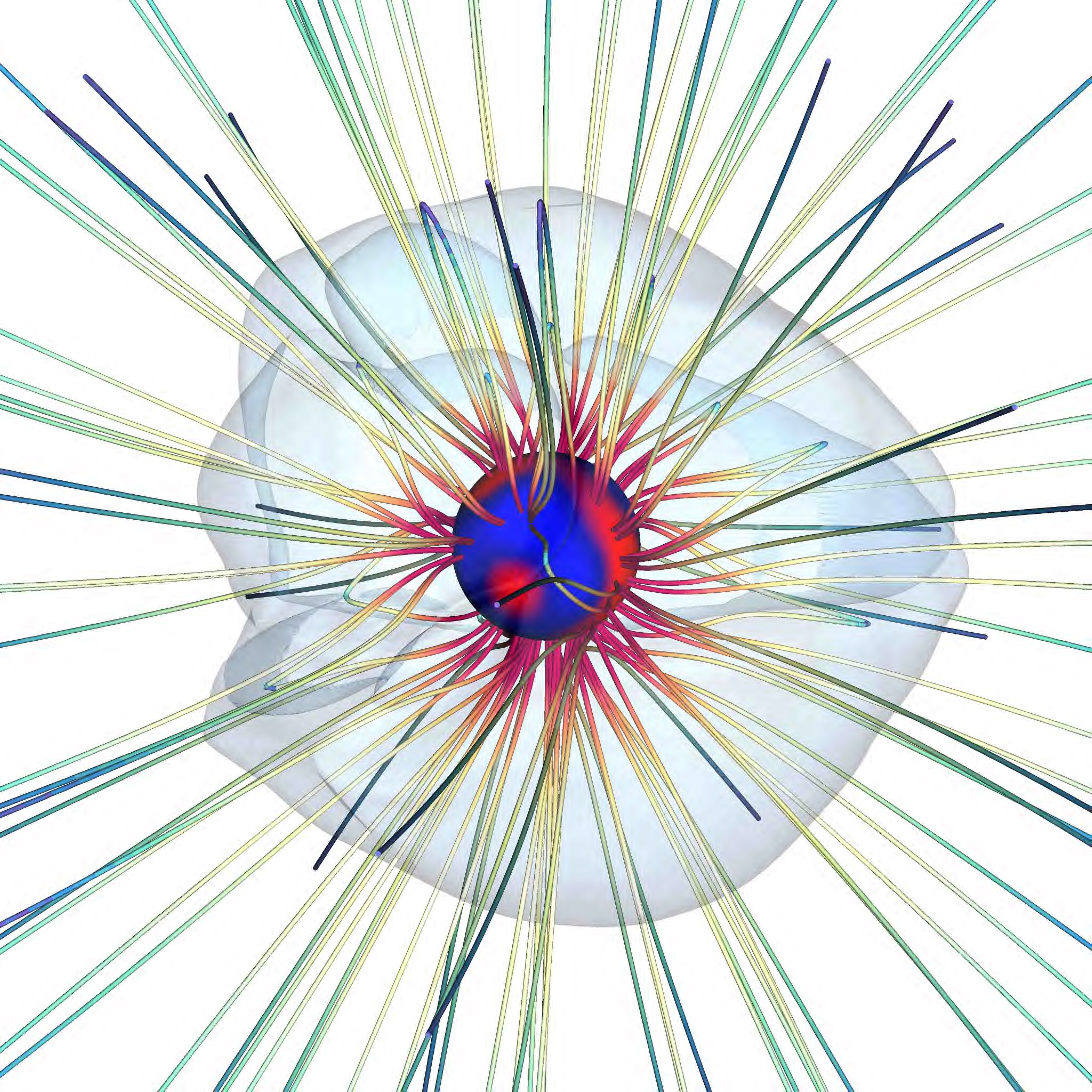}
  \includegraphics[width=0.4\linewidth,angle=90]{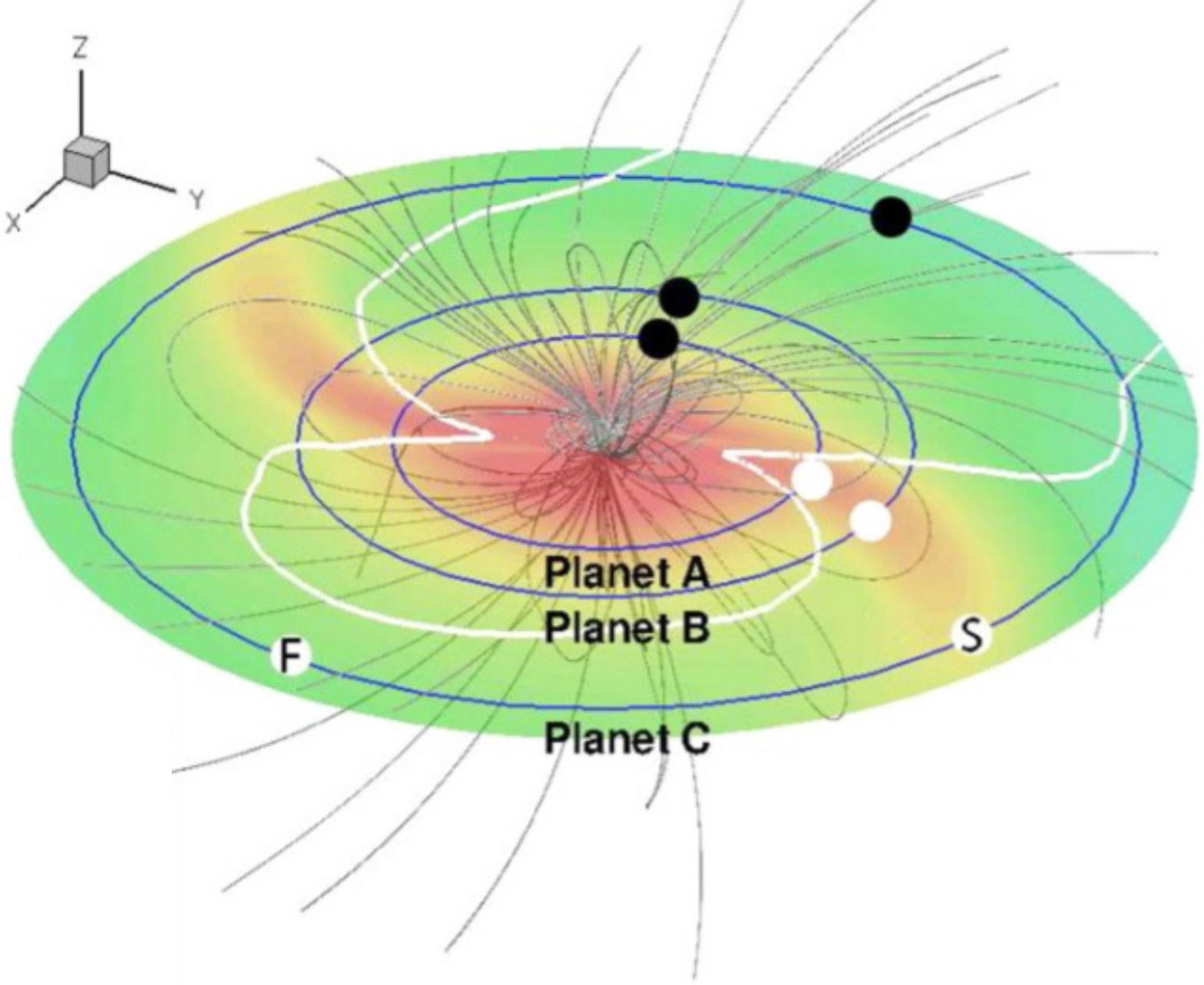}
  \caption{\hbindex{Stellar wind} models of HD 189733 (left,
    see \citealt{Strugarek:2014uv}) and EV Lac (right,
    adapted from \citealt{Cohen:2014eb}) based on the spectropolarimetric
    reconstructions of their large scale magnetic field. Reproduced with
    the permission of AAS.}
  \label{fig:stellar_wind_models}
\end{figure}

Early work on \hbindex{Alfv\'en wings} in exoplanetary systems were carried out
by \citet{Preusse:2005cl,Preusse:2006iu}, where they used a
Weber-Davis \hbindex{stellar wind} model \citep{Weber:1967kx} to estimate the amount of energy that
could be channeled by the magnetic interaction for a planet-size
obstacle. This estimation was recently revisited by
\citet{Saur:2013dc} with many more exoplanets and a more
sophisticated \hbindex{Alfv\'en wings} model. Strong \hbindex{planet
migration} due to magnetic torques around T Tauri stars
\citep{Lovelace:2008bl} and proto-stars \citep{Bouvier:2015kq} were
also assessed using similar \hbindex{stellar wind} models. Observed stellar
magnetic fields \citep[see \textit{e.g.}][]{Donati:2009if} allow to
model more realistically \hbindex{stellar winds}, and as a result provide more
quantitative estimates of \hbindex{SPMI}s (for
more details see \citealt{Vidotto:2017wf}, \citealt{Moutou:2017vx}, Figure
\ref{fig:stellar_wind_models}, and \citealt{Vidotto:2014kk,Cohen:2014eb,Llama:2013il,Strugarek:2014uv, AlvaradoGomez:2016il,AlvaradoGomez:2016iv}).



In all these studies,
simplified models of the
interaction with the planet 
are assumed (either analytical models, or localized simulations
at particular orbital phases using the plasma conditions from the
\hbindex{wind} model as in \citealt{Cohen:2014eb,AlvaradoGomez:2016iv}). We will see in the following
sections that ultimately, both a realistic \hbindex{stellar wind} and a realistic model of
the star-wind-planet coupling need to be considered to quantitatively
model the effects of \hbindex{SPMI}s.


\subsection{Planet properties and star-planet magnetic
  interactions}
\label{sec:plan-char-star}

The magnetic properties of the orbiting planet also change qualitatively the
\hbindex{SPMI} (as one would expect), and two families of interaction can be identified
\citep[following, \textit{e.g.}][]{Zarka:2007fo}:

\begin{itemize}
\item \textbf{Unipolar interaction}: weakly/not magnetized planet in
  a magnetized wind
\item \textbf{Dipolar interaction}: magnetized planet in a magnetized wind
\end{itemize}

One important distinction between the two families lies in the
possibility of magnetic reconnections in the case of dipolar
interaction, whereas in the case of unipolar interaction the \hbindex{stellar
wind} magnetic field generally permeates into some parts of the planet
without necessarily reconnecting. 
Whether or
not a close-in planet is able to sustain its own magnetosphere 
is out of the scope of this review 
(
see \citealt{Stanley:2010ba,Jones:2011fx}). 
 We will discuss various
modelling efforts of the two interaction types in the next section, we
now focus on giving a general overview of them.

\begin{figure}[!htbp]
  \centering
  \includegraphics[width=\linewidth]{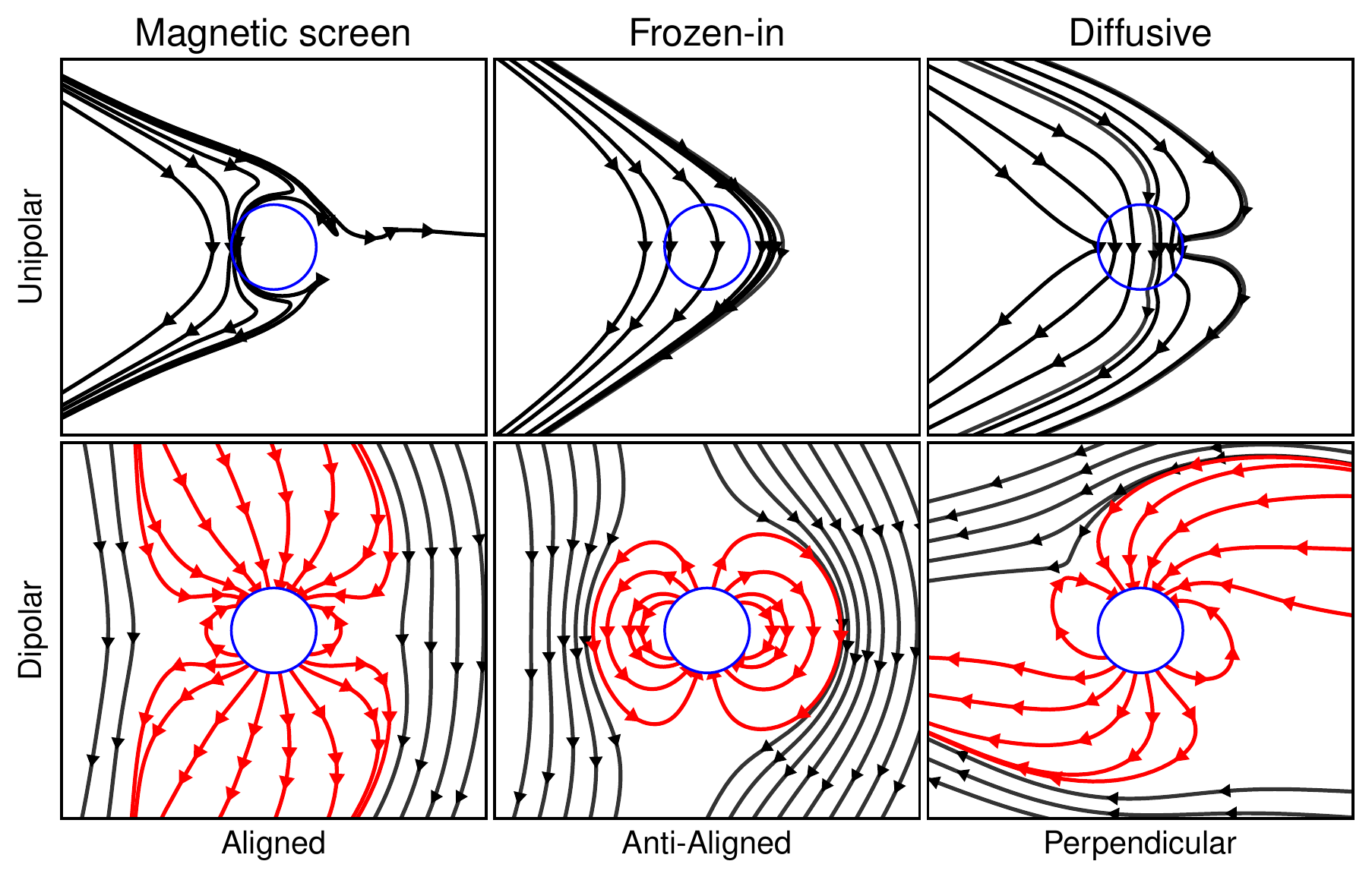}
  \caption{Magnetic interaction for different topologies. The planet
    is indicated in blue, the wind magnetic field lines in black. The upper
    row shows three cases of the unipolar interaction scenario
    (magnetic screening, magnetic field drag by the orbiting planet,
    ohmic dissipation inside the planetary body), and the lower row
    three cases of the dipolar interaction scenario (aligned configuration,
    anti-aligned configuration --closed magnetosphere--, perpendicular
    configuration). In the dipolar configuration the magnetic field
    lines connected to the planetary field are shown in red. The
    magnetic configurations were taken from numerical
    models published in \citet{Strugarek:2014gr,Strugarek:2015cm}.}
  \label{fig:unipolar_vs_dipolar}
\end{figure}

\runinhead{Unipolar interaction} The unipolar interaction occurs when
the magnetic field of the planet can be neglected compared to the
\hbindex{stellar wind} magnetic field. Several cases of unipolar interaction
need to be distinguished (upper panels in Figure
\ref{fig:unipolar_vs_dipolar}), depending on the ionization and the
resistivity of the planet material
\citep{Laine:2008dx,Laine:2012jt}. If the planet material is not or
weakly ionized, the magnetic field inside the planet is only subject
to ohmic dissipation and two extreme cases are identified: if the resistivity
inside the planet sufficiently high, the \hbindex{stellar wind} magnetic field
penetrates only on a small skin depth inside the planet, while in the
opposite case the magnetic diffusivity is low and the \hbindex{wind} magnetic
field permeates into the whole planetary volume
\citep{Laine:2008dx}. If the planet material is ionized, induction can
occur inside the planet and the situation is slightly more complex:
different regimes of the interaction occur depending on the ratio
between the advection across the planetary obstacle and the ohmic
dissipation time-scale inside the planet. In this latter case, the
extreme situations are realized when the planet is able to completely
drag the magnetic field lines along its orbital motion (similar to a
\textit{frozen-in} situation of ideal MHD, see also
\citealt{Laine:2012jt,Strugarek:2014gr}), and when the planet
effectively screens the surrounding wind magnetic field, leaving a
magnetic cavity in the planetary interior. It must be noted that in
reality, more complicated situations occur with strong anisotropies
in the conductive properties of the planet material, due to
\textit{e.g.} day-night asymmetries that are likely realized in
tidally-locked states for close-in exoplanets.

\runinhead{Dipolar interaction} In the dipolar case the interaction
occurs between the \hbindex{stellar wind} magnetic field and the planetary
magnetosphere. Here the critical parameter of the interaction is the
topology of the interaction (three topologies are illustrated in the
lower panels of Figure \ref{fig:unipolar_vs_dipolar}) that
determines the location of the reconnection sites between the two
fields (we assume for the discussion here that there is no \hbindex{shock} at
the nose of the magnetosphere). If the planetary field is locally
aligned with the \hbindex{stellar wind} field, reconnections occur on the
(magnetic) equatorial plane and the polar field lines are directly
connected to the \hbindex{stellar wind}. This is the so-called \textit{open
  magnetosphere} case, where only a small volume of closed planetary
field lines exists around the equatorial plane. In the anti-aligned
configuration, the \textit{closed magnetosphere} case is realized and
the reconnection sites are located near the polar caps of the
planetary magnetosphere. We immediately see here that the size of the
interacting obstacle drastically changes with such a change of
topology, and consequently we expect the strength of the \hbindex{SPMI}s to
strongly vary with the topology of the interaction (we will quantify
these aspects in the next sections). 

\subsection{Connection to planet-satellite interactions}
\label{sec:conn-plan-satell}

\begin{figure}[!htbp]
  \centering
  \includegraphics[width=0.39\linewidth]{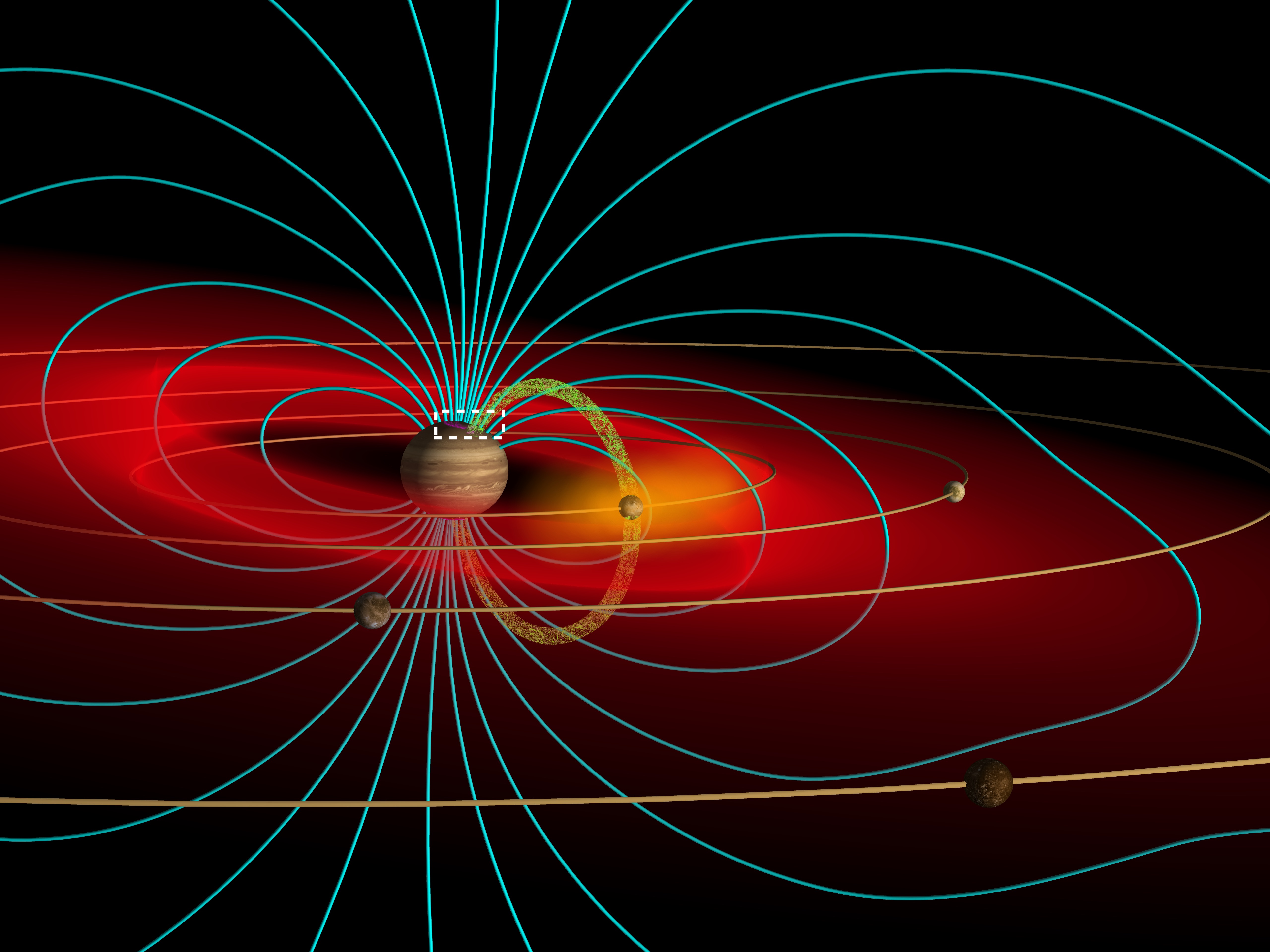}
  \includegraphics[width=0.6\linewidth]{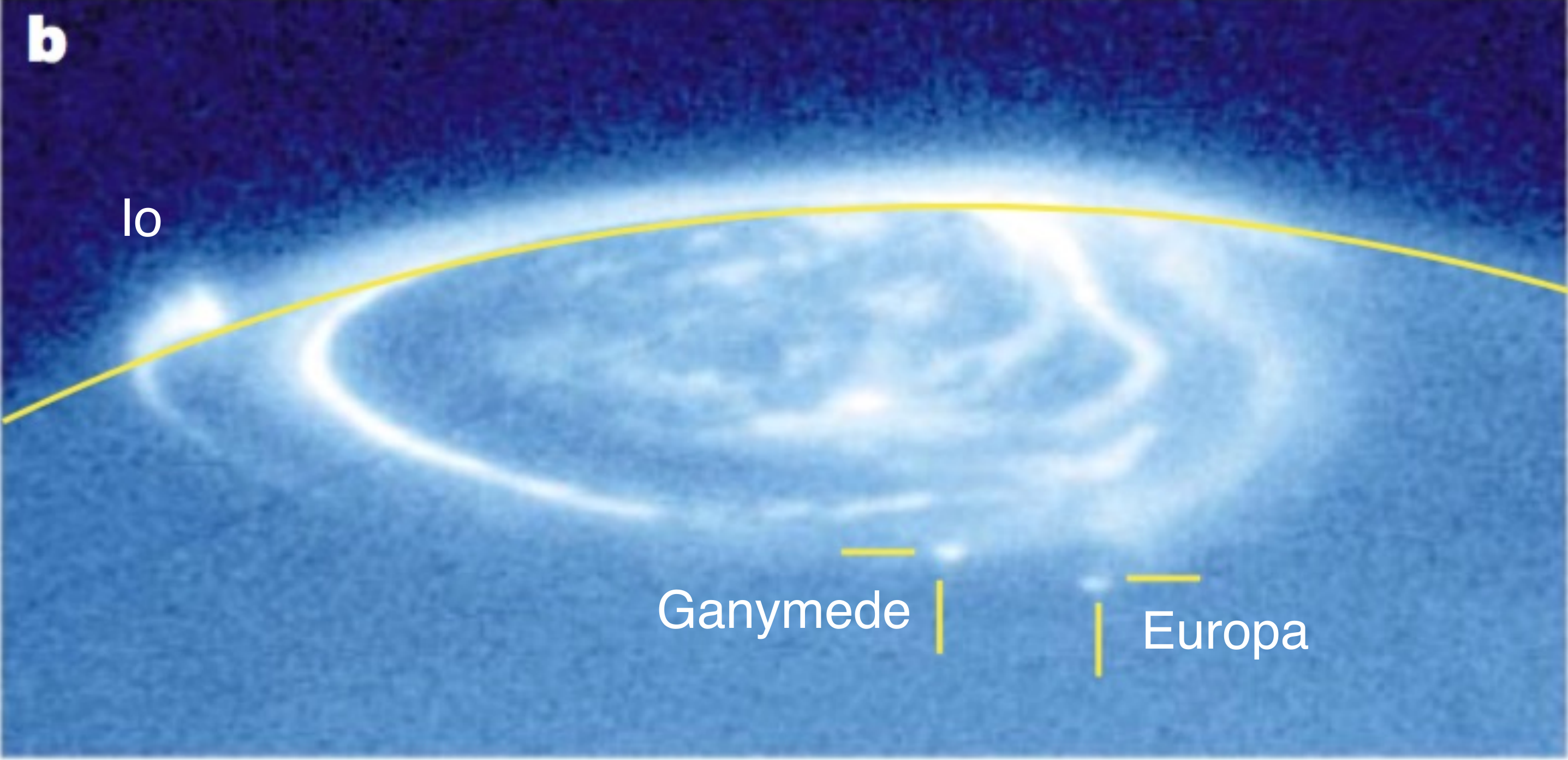}
  \caption{\textit{Left} Artist image of the jovian system, where the
    Io torus can be seen in red/orange and the magnetic connection to
    jupiter is traced by the green tube. Adapted from a rendering from
    J. Spencer
    \url{http://www.boulder.swri.edu/\%7Espencer/digipics.html}. 
\textit{Right} Aurorae at the
    pole of Jupiter triggered by the magnetic
    interactions with the jovian satellites in the magnetosphere
    (adapted from \citealt{Clarke:2002tx}). Reproduced with
    the permission of AAS.}
  \label{fig:jovian_aurorae}
\end{figure}

\hbindex{SPMI}s bear strong similarities with the interaction of a natural
satellite with its hosting planet magnetosphere. As a result, most of
the concepts presented in this review take their roots in pioneering
studies of Io, Ganymede, and other jovian and cronian moons (see
Figure \ref{fig:jovian_aurorae}). The main
difference here is that close-in planets orbit in a more dynamical medium,
the \hbindex{stellar wind}, and are exposed to extreme eruptive events
triggered by the stellar magnetism. A more in depth parallel between
close-in star-planet systems and satellite-planet systems may be found
in \citet{Neubauer:1998bw,Zarka:2007fo,Saur:2013dc}.

\section{Models of the various cases of star-planet magnetic interactions}
\label{sec:various-cases-star}

We now detail recent efforts in modelling \hbindex{SPMI}s in close-in
star-planet systems. We discuss successively the unipolar and dipolar
interaction cases.

\subsection{The unipolar interaction case}
\label{sec:unip-induct-model}

\subsubsection{Analytical considerations}
\label{sec:analyt-cons}

\begin{figure}[!htbp]
  \centering
  \includegraphics[width=0.6\linewidth]{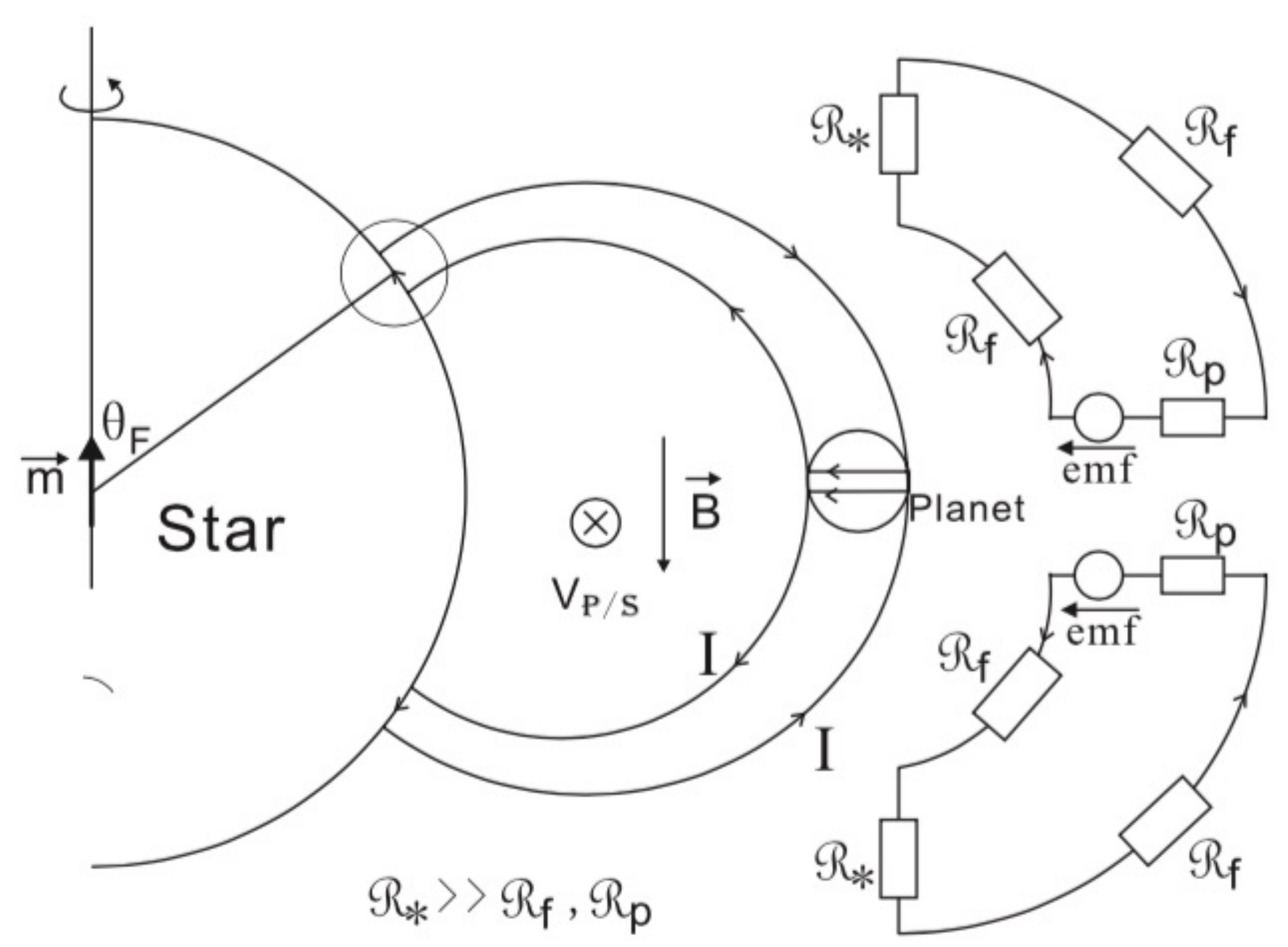}
  \includegraphics[width=0.39\linewidth]{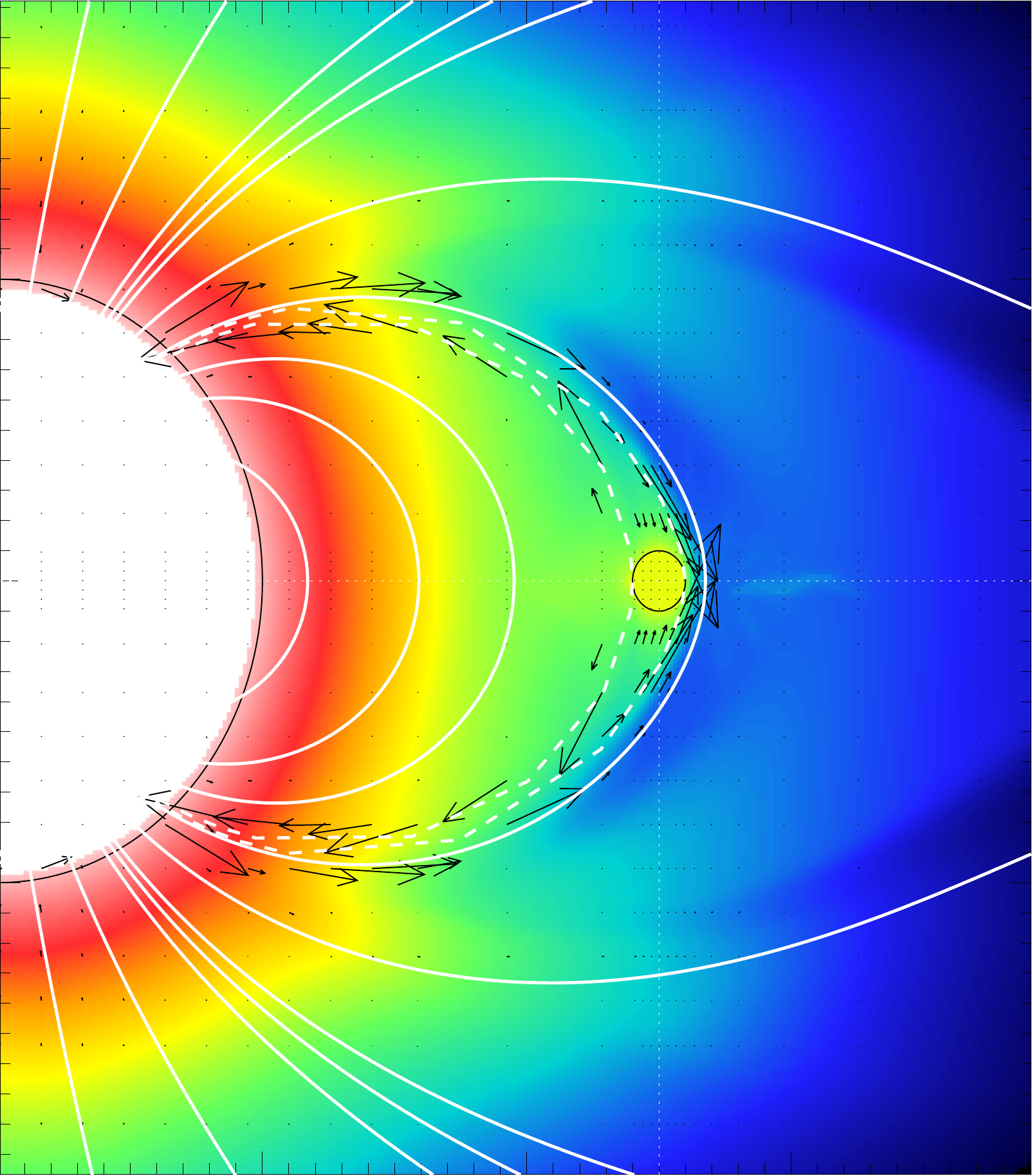}
  \caption{\textit{Left.} Schematics of the unipolar inductor model, adapted from
    \citet{Laine:2012jt}. The star-planet system is approximated by
    an electric circuit composed of an electro-motive force (from the
    orbital motion of the planet) and several resistances (the planet,
    the \hbindex{stellar wind}, the stellar surface). In this model, it is assumed
    that the alfv\'enic perturbations are fast enough to travel back and forth
    between the planet and the star. \textit{Right.} Numerical
    simulation of the unipolar inductor model from
    \citet{Strugarek:2012th,Strugarek:2014gr}. The current system that
    self-consistently develops is shown by the black arrows, the
    magnetic field lines are shown in white, and the logarithmic colormap traces
    the plasma density.}
  \label{fig:unipolar_cases}
\end{figure}

The unipolar interaction case has been analytically studied in an
extensive work published in \citet{Laine:2008dx,Laine:2012jt}. In
their model, an analogy is drawn between the current circuit that
develops in \hbindex{SPMI}s and a standard electric circuit (see Figure
\ref{fig:unipolar_cases}) composed of four
resistances (planet, interplanetary medium, star, interplanetary
medium) for each \hbindex{Alfv\'en wing}, and a generator (the planet
differential motion with the ambient wind). The model is
considered to be valid as long as the circuit is closed, which the
authors choose to close at the stellar surface (note that the circuit
could close along the path of the wing due to the reflection of the
waves inside the Alfv\'en wings themselves, see \textit{e.g.}
\citealt{Neubauer:1998bw}). It means that the following expressions
are valid as long as the Alfv\'en waves have the time to travel back
and forth between the planet and the star while the planet continues
its orbital motion. We saw in the previous sections that this was likely
not the case for solar twins. In the context of TTauri stars
on which the authors focused, the stellar field (and the Alfv\'en
speed) is orders of magnitude larger, which ensures this condition
(Equation \ref{eq:travel_back_forth}) can
be easily satisfied. Based on the electric circuit analogy, it was
shown that the total torque applied to such close-in planets in the unipolar
interaction case could be written 
\begin{equation}
  \label{eq:torque_unipolar_LL}
  \mathcal{T} \propto R_{\rm orb}^{-4} \left(\omega_P - \omega_\star
  \right) \,
\end{equation}
where $\omega_P$ is the orbital frequency and $\omega_\star$ the
rotation frequency of the star (we simplified the original expression
of \citealt{Laine:2012jt} for a
circular orbit) . This estimate gives a \hbindex{migration} time-scale 
\begin{equation}
  \label{eq:taup_unipolar_LL}
  \tau_P = \frac{J_P}{2\mathcal{T}} \propto R_{\rm orb}^{6}\, ,
\end{equation}
where $J_P=M_P\left(GM_\star R_{\rm orb}\right)^{1/2}$ is the orbital
angular momentum of the planet.

Using this unipolar model, it is then straightforward to show
that close-in planets under orbital periods of about three days
migrate due to the \hbindex{SPMI} on a time-scale of the order of a few million
years when the hosting star is a standard TTauri star. We remind the
reader here that these estimates rely on several debatable hypotheses,
among which the internal resistivity profile of the planet, that
enters the proportionality factor in Eq. \ref{eq:torque_unipolar_LL}, is highly
uncertain today.

\subsubsection{Numerical models}
\label{sec:numerical-models}

Very few numerical studies of the unipolar interaction case have been
carried out as of today. One notable exception was the work published
in \citet{Strugarek:2014gr} using a reduced 2.5D (axisymmetric)
geometry (see Figure \ref{fig:unipolar_cases}). In this model, the planet was considered to be fully ionized
and as a result was able to drag the \hbindex{stellar wind} magnetic field along
its orbit. By varying the magnetic diffusivity inside the planet,
\citet{Strugarek:2014fr} showed various cases of unipolar interaction
(see also Figure \ref{fig:unipolar_vs_dipolar})
from the creation of a magnetic cavity when an ionospheric layer exists
in the planet atmosphere, to cases where the \hbindex{stellar wind} magnetic
field is either primarily dragged or dissipated by the planet (similar
to the case modelled with the analytical approach presented in the
previous section). 

The former case was extensively characterized in \citet{Strugarek:2014gr} in
a parameter space exploration by varying the orbital radius of the
close-in planet. The magnetic torque applied to the planet was
systematically assessed, and the associated \hbindex{migration} time-scale
was found to be proportional to $R_{\rm orb}^{5.5}$. In spite
of the geometry approximation embedded in the numerical model, these
results compare well with the analytical estimations giving
$\tau_P\propto R_{\rm orb}^{6}$ (Equation \ref{eq:torque}). Three-dimensional simulations
are now required to further refine the parametrization of 
the magnetic torque in the unipolar case. Such simulations would also
help to better characterize the other unipolar interaction cases
that have not today been satisfactorily modelled (see above). They are
notably needed to assess how the current system that systematically
develops can be closed in either the wind itself or in the stellar (sub-)surface layers.

\subsection{The dipolar interaction case}
\label{sec:dipolar-model}

The dipolar interaction bears some ressemblance with the Earth-solar
wind interaction, and as a result has received more attention so
far. Once again, we detail first the analytical approaches to model this case,
and then report on recent numerical developments.

\subsubsection{Analytical considerations}
\label{sec:analyt-cons-2}

Two main analytical approaches have been followed in the literature so
far. In both models, the topology of the interaction (relative
direction of the planetary field compared to the ambient \hbindex{stellar wind}
magnetic field) determines the strength of the interaction, with two
extreme cases of aligned and anti-aligned configurations (see also
bottom panels in Figure \ref{fig:unipolar_vs_dipolar}). These two
models can be summarized as follows.

A description based on a magnetohydrostatic equilibrium, characterized
by force-free magnetic fields, has been carried out by
\citet{Lanza:2008fn,Lanza:2009fp,Lanza:2012jv,Lanza:2013gj}. In this
model, the orbital motion of the planet stresses the ambient field and
the excess energy (compared to the planet-free, force-free field) is
stored in the flux tube connecting the planet to the star. As
dissipation sets in, a saturated steady state can be reached and the
\hbindex{Poynting flux} through the flux tube can be analytically shown to reach
values of $10^{20}-10^{21}$ W for close-in planets around typical
solar-type stars \citep{Lanza:2013gj,Lanza:2015uu}.

\begin{figure}[!htbp]
  \centering
  \includegraphics[width=0.4\linewidth]{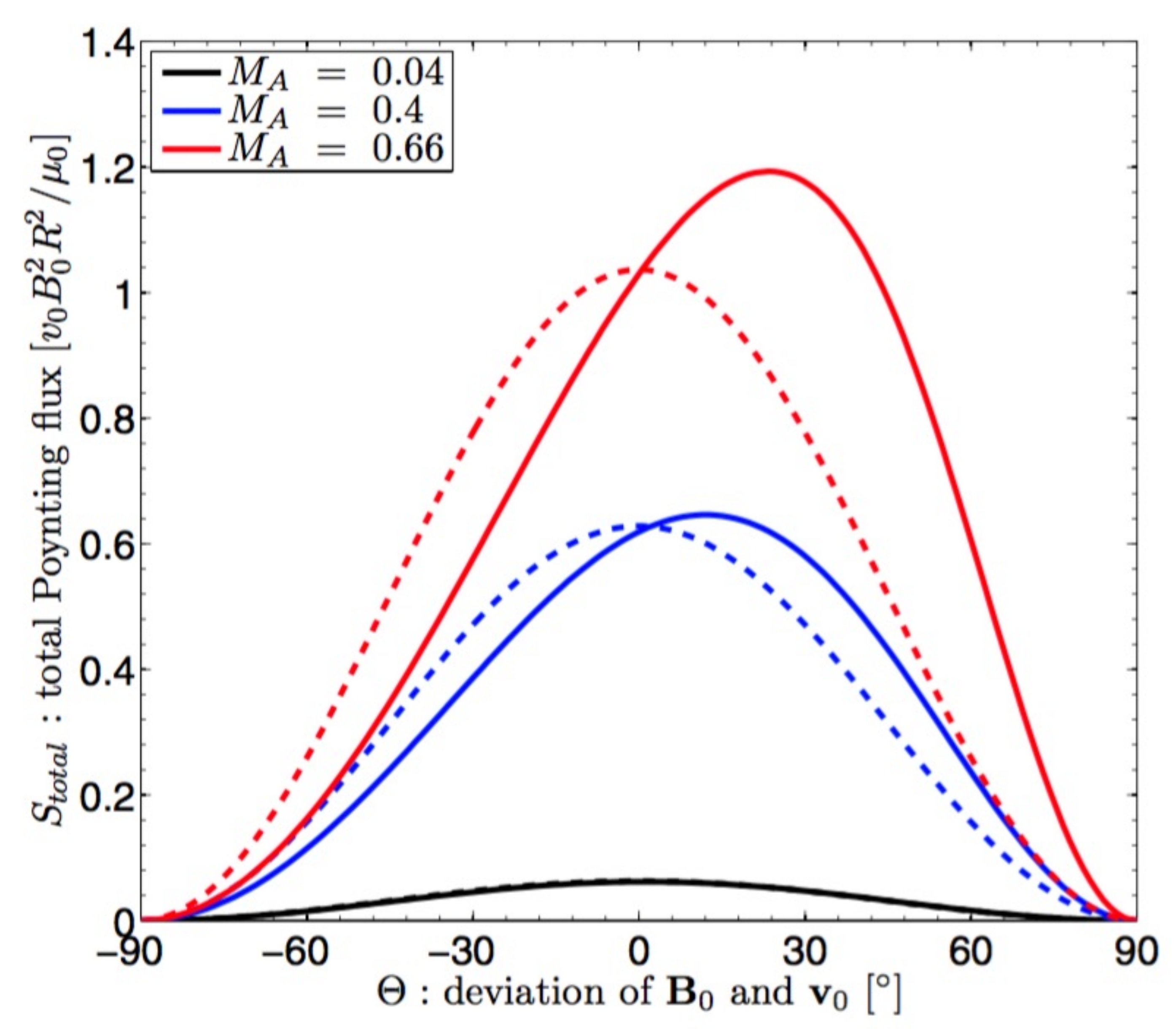}
  \includegraphics[width=0.4\linewidth]{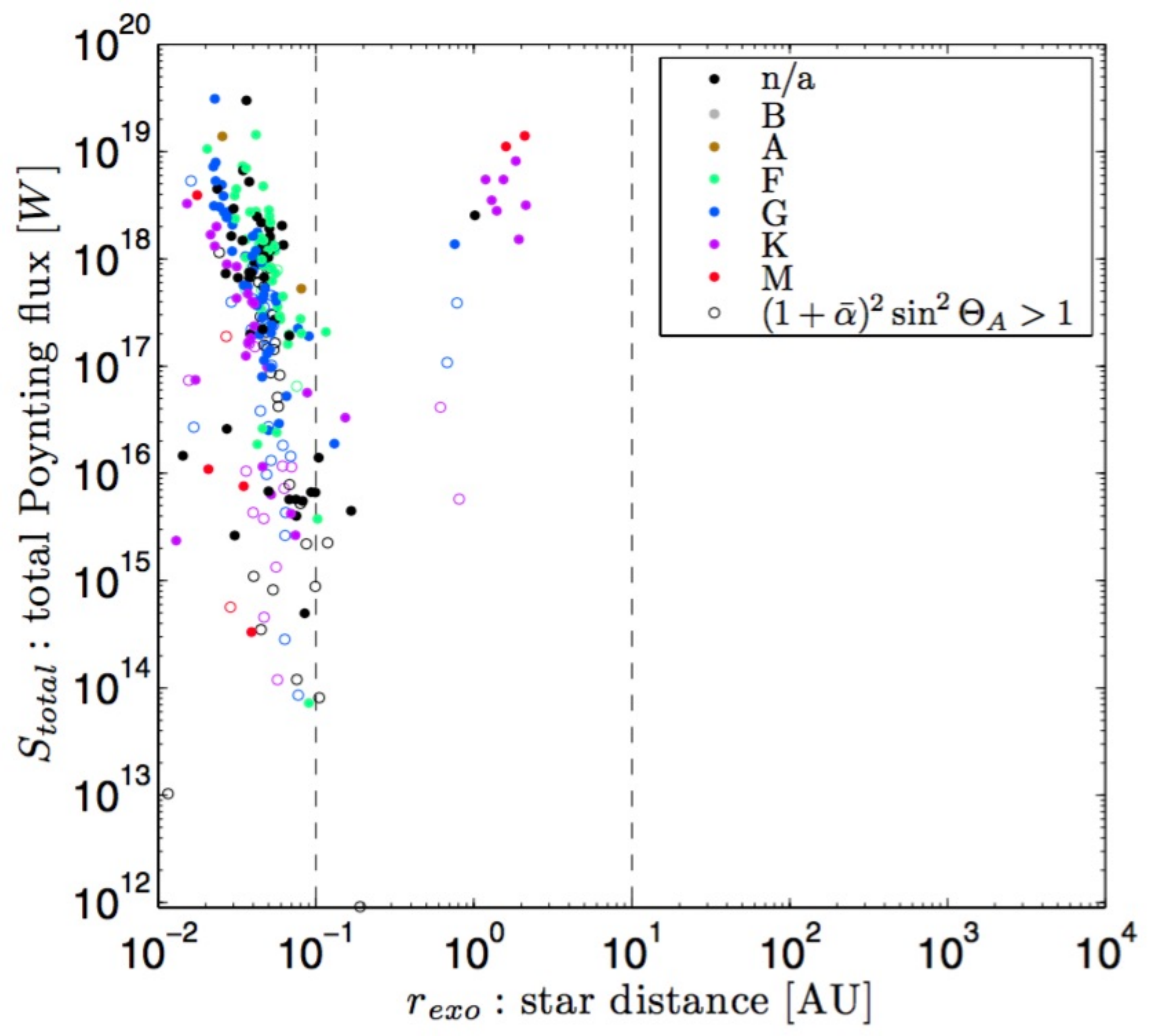}
  \caption{\textit{Left} \hbindex{Poynting flux} as a function of
    topology ($\Theta$) and
    Alfv\'en Mach number $M_a$. \textit{Right} Estimate of the
    \hbindex{Alfv\'en wing} \hbindex{Poynting flux} for detected exoplanets (as of
    2013). Both estimates are based on the analytical model of
    \citet{Saur:2013dc}, from were the figures were adapted.}
  \label{fig:dipolar_analytical}
\end{figure}

The other family of models is based on the analogy with
planet-satellite interactions and the concept of \hbindex{Alfv\'en wings}. It is
a more general class of models, as it does not suppose any particular
configuration (\textit{e.g.} potential) for the \hbindex{stellar wind} magnetic field.
The \hbindex{Alfv\'en wings} current system has been clearly laid out in 
\citet{Saur:2013dc}. 
They found that the Alfv\'en
wing cross-section $R_{\rm aw}$ varies such that
\begin{equation}
  \label{eq:aw_cc}
  R_{\rm aw} \propto \sqrt{\cos \left(\frac{\Theta_M}{2}\right)}\, ,
\end{equation}
where $\Theta_M$ is the relative angle between the planetary (dipolar)
magnetic field and the ambient \hbindex{stellar wind} direction at the
planet position. As
a result, they found analytically that the \hbindex{Alfv\'en wings} are
vanishingly small in the anti-aligned configuration ($\Theta_M=\pi$),
and are maximized in the aligned configuration ($\Theta_M=0$). In the
former case, in reality, some magnetic reconnection still occurs
between the wind and the planetary magnetosphere, leading to very
diminished \hbindex{Alfv\'en wings}-like structure (see hereafter). \\
The energetics of the interaction were also shown to depend on the
ratio between the Alfv\'en conductance $\Sigma_A$ and the (integrated)
Pedersen conductance of
the ionosphere ($\Sigma_P$) through the coefficient \citep{Neubauer:1998bw,Saur:1999kh}
\begin{equation}
  \label{eq:alpha_ratio}
  \bar{\alpha} = \left(1+2\frac{\Sigma_A}{\Sigma_P}\right)^{-2}\, ,
\end{equation}
knowing that within the \hbindex{Alfv\'en wing} model the Alfv\'en
conductance is defined (in Gaussian units) by
\begin{equation}
  \label{eq:alfven_conductance}
  \Sigma_A= \frac{c^2 M_a}{4\pi v_o\left(1+M_a^2-2M_a\cos\Theta\right)^{1/2}}\, ,
\end{equation}
where $\Theta$ is the angle between the relative speed ${\bf v}_o$ and
the wind magnetic field ${\bf B}_w$. The efficiency coefficient
$\bar{\alpha}$ is maximized
when the Pedersen conductance is large compared to the Alfv\'en
conductance. For planets in close-in orbit around solar-type stars, $\Sigma_A$ can reach
values of a few $10^{12}$ cm/s \citep{Strugarek:2016ee}. Based on
estimates from solar-system planets and moons, the Pedersen
conductance is expected to be of the order of $10^{13}$ cm/s
\citep[see Figure \ref{fig:dipolar_analytical} and][]{Saur:2013dc}. As
of today, $\bar{\alpha}$ has thus been
neglected in applications to close-in exoplanets. Finally, using their
analytical \hbindex{Alfv\'en wing} model and simple Parker-like wind solution,
Saur \textit{et al.} estimated the \hbindex{Poynting flux} through the wings for the
close-in exoplanets
known in 2013. They found that the Poynting could vary from $10^{14}$
to a few $10^{19}$ W depending on the orbital distance. Nevertheless,
no observational constraints were available to characterize properly the
wind of the planet-hosting stars, hence these estimates have to be taken with caution.

In real
exoplanetary systems, 
the interaction state is likely to vary on short
(inhomogeneities along the planetary orbit)
and large (reversals of the stellar and/or planetary
magnetic fields) time-scales. Large uncertainties on the \hbindex{stellar wind}
characteristics of the central star and large uncertainties on the ionospheric
properties of the exoplanet (value of the efficiency parameter
$\bar{\alpha}$) also make the quantitative estimates of \hbindex{SPMI}s
perilous. Numerical simulations can be used to tackle some of
these aspects, which we now turn to.

\subsubsection{Numerical models}
\label{sec:numerical-models-2}

Numerical models of star-planet interactions take their roots in the
pioneering work of \citet{Ip:2004ba} who first simulated the local
interaction of a close-in giant exoplanet with the ambient \hbindex{stellar
wind} plasma. Nevertheless, 
\hbindex{SPMI}s depend on the plasma characteristics from the base of the
stellar corona to the vicinity of the planet and on the planet
magnetic configuration. 

Global simulations were developed for the first time by
\citet{Cohen:2009ky,Cohen:2010jm,Cohen:2011gg}, in which advanced,
solar-calibrated \hbindex{stellar wind} simulations
were adapted to include a close-in orbiting planet (treated as a boundary
condition). In their later work, they introduced a boundary condition
inside the simulation domain moving with time to follow the orbital
path of the planet. These early simulations strikingly showed the
natural time-variability one expects from \hbindex{SPMI}s, as well as possibly
very large lags between the orbital
phase of the planet and the stellar subpoint where the magnetic
interaction connects. Hybrid approaches for which the \hbindex{stellar wind} and
the planet vicinity are modelled separately for the same system have also been
carried out by \citet{Kopp:2011if,Cohen:2014eb,Cohen:2015gd,AlvaradoGomez:2016il,AlvaradoGomez:2016iv}. 

\begin{figure}[!htbp]
  \centering
  \includegraphics[width=0.49\linewidth]{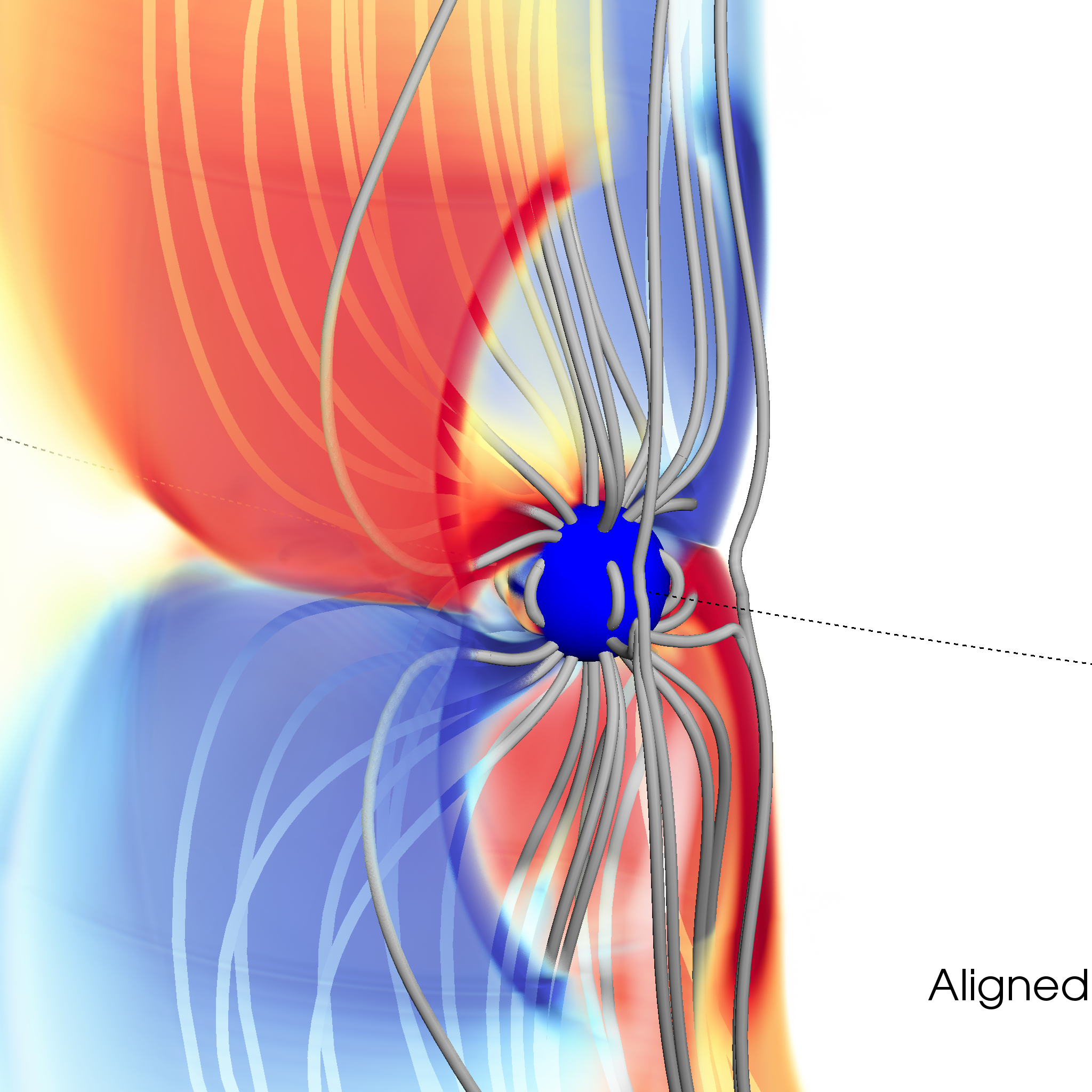}
  \includegraphics[width=0.49\linewidth]{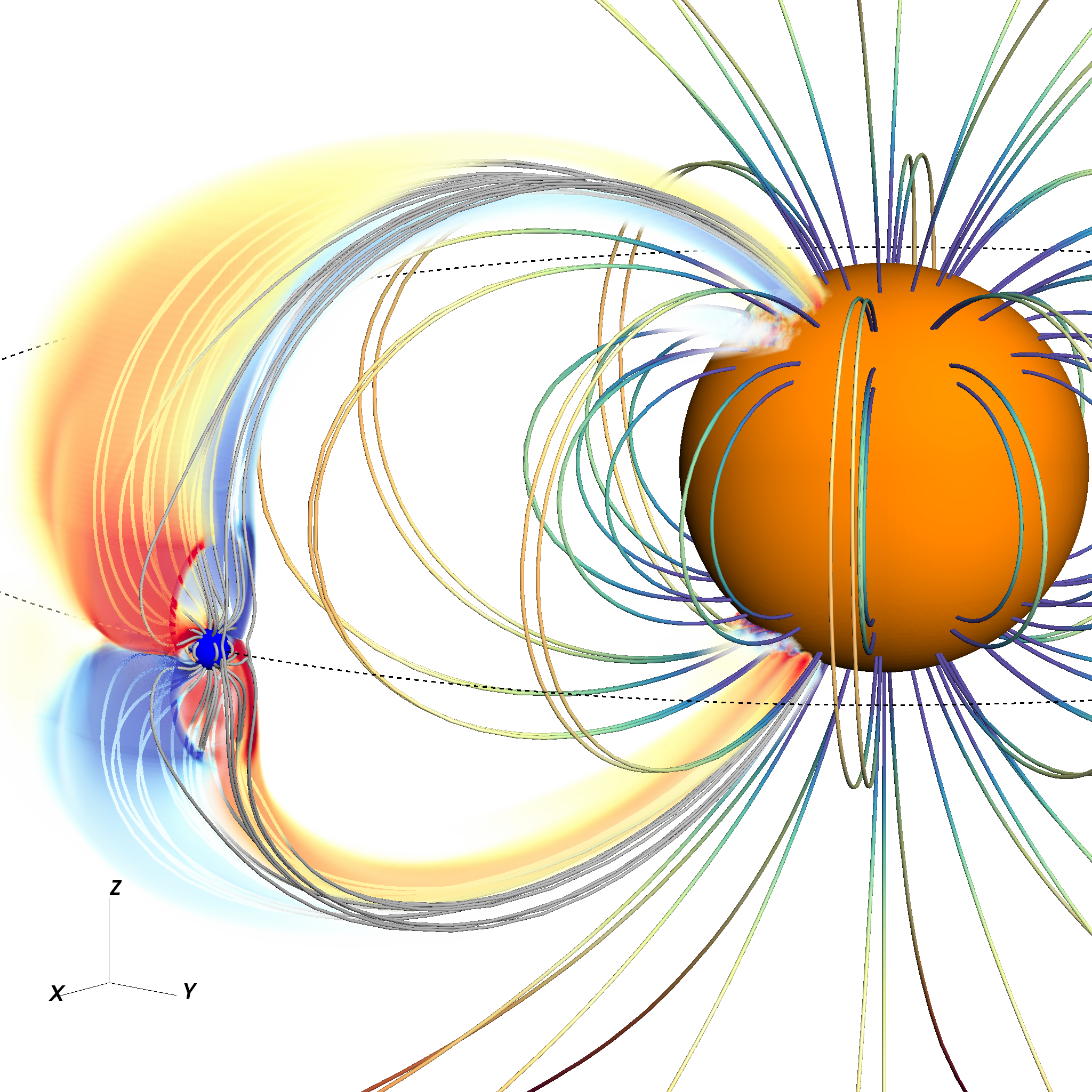}
  \includegraphics[width=0.49\linewidth]{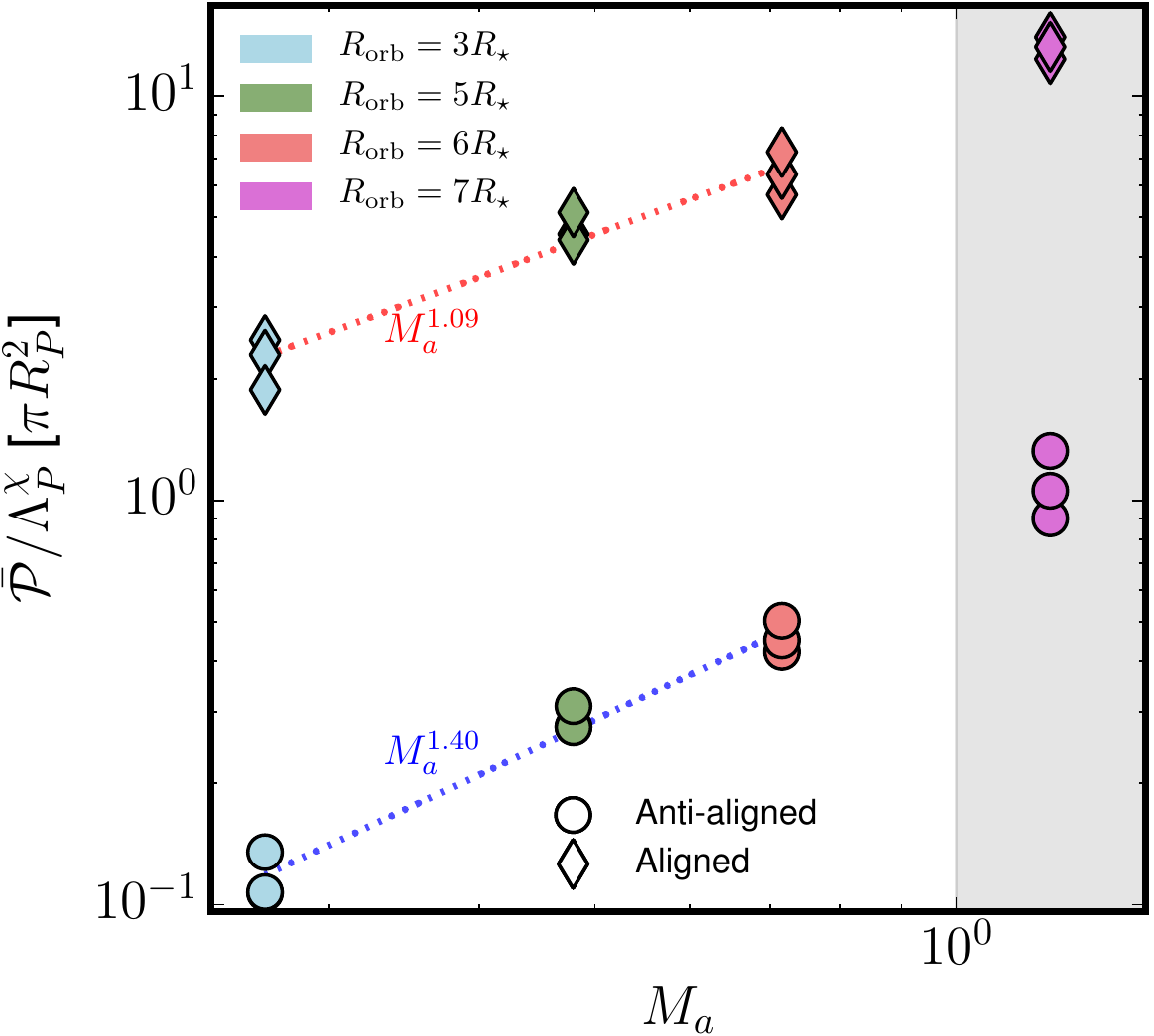}\hfill
  \includegraphics[width=0.48\linewidth]{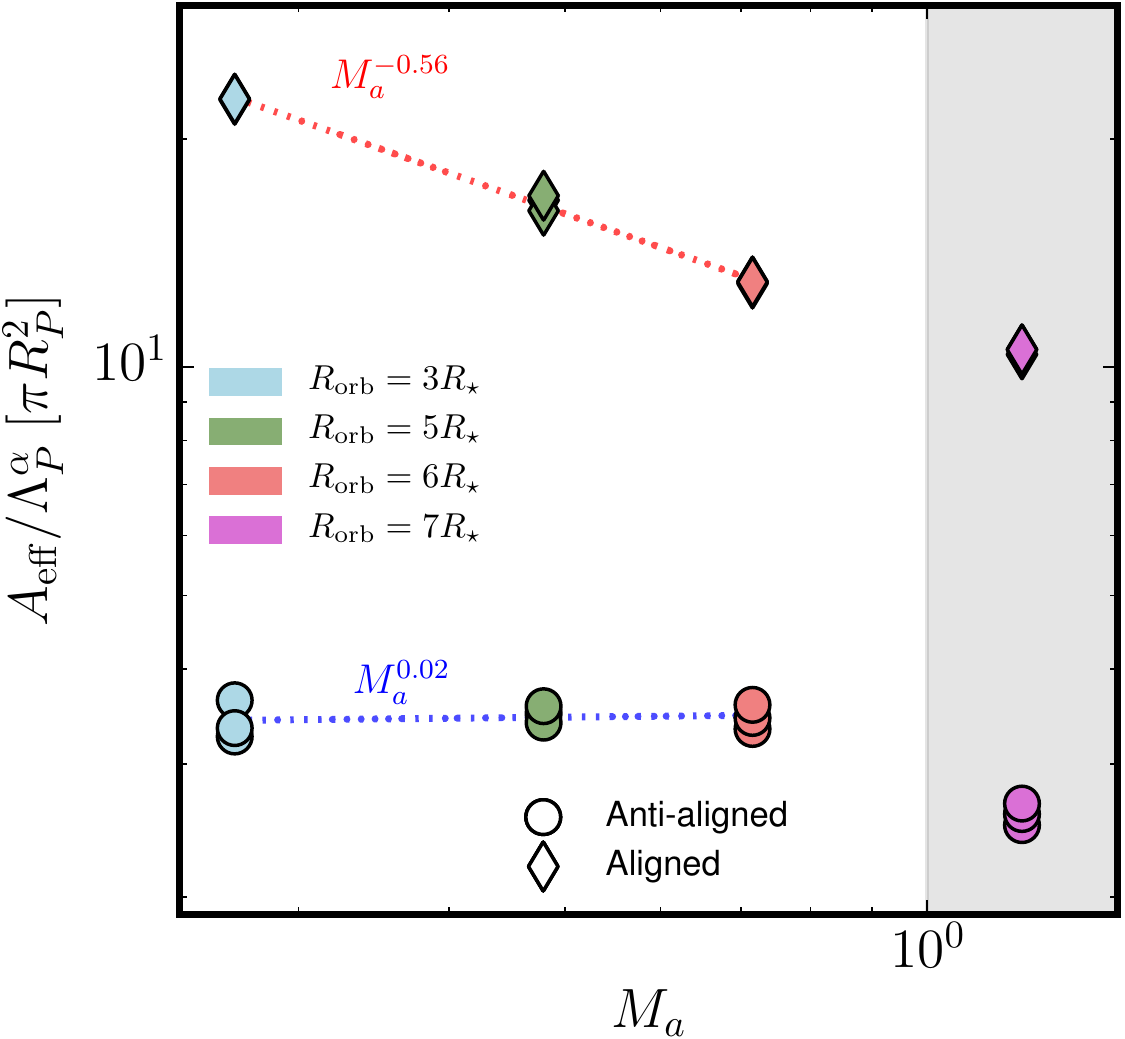}
  \caption{\textit{Top panels} 3D numerical model of 
    \hbindex{Alfv\'en wings} in the aligned configuration. The coloured volumes in blue/red trace the currents
    underlying the wings (the volume is cropped to make the planet
    apparent). The stellar magnetic field lines are colour-coded by
    the magnetic field strength, and the planetary field is shown by
    the grey tubes. Adapted from \citet{Strugarek:2015cm}. \textit{Bottom panels} Scaling-laws of the
    \hbindex{Poynting flux} in one \hbindex{Alfv\'en wing} (left) and of the torque applied to
    the planet (right) deduced from the numerical model of
    \hbindex{SPMI}s of \citet{Strugarek:2015cm}. Both are shown as a function of the alfv\'enic Mach number
    $M_a$. Adapted from \citet{Strugarek:2016ee}. Reproduced with
    the permission of AAS.}
  \label{fig:dipolar_numerical}
\end{figure}

Building on the 2.5D approximation used in \citet{Strugarek:2014gr},
\citet{Strugarek:2015cm} also developed a global 3D model incorporating
both the orbiting planet and the star, using as a
first approach simple axisymmetric magnetic
configurations (see top panels in Figure \ref{fig:dipolar_numerical}). A special effort was carried out to develop adequate
boundary condition for both the star and the planet, which are
critical to correctly quantify \hbindex{SPMI}s. The
\hbindex{stellar wind} boundary condition consists in a three-layer boundary
condition, ensuring accurate conservation properties throughout the
simulated \hbindex{stellar wind}
\citep{Zanni:2009kc,Strugarek:2014fr,Strugarek:2014tf}. The boundary
condition at the planet is defined by a buffer layer in which only the
magnetic field of the planet is allowed to change, mimicking crudely
a thick ionospheric layer \citep[similarly to the approach developed in][in the context of
Ganymede in the jovian system]{Jia:2009ha}. 
More
advanced planetary boundary conditions have been
developed in the recent years. \citet{Duling:2014en} developed a
generic boundary condition for non-conductive planetary bodies, and
showed that the choice of boundary condition for the planet has
a drastic impact on the development of the magnetic interaction (see,
\textit{e.g.}, the change in the current system on their
Fig. 3). Ultimately, the interaction of a planetary magnetosphere with
the wind of its star could be modelled with a higher precision using
a magnetosphere-ionosphere coupling boundary condition, as already
developed for simulations of the magnetosphere of the Earth 
\citep[\textit{e.g.}][]{Goodman:1995hf,Merkin:2010dr}. 

The simple geometry used in \citet{Strugarek:2015cm} allowed a quantitative comparison
between the numerically modelled \hbindex{Alfv\'en wings} and the analytical
work of \citet{Saur:2013dc}. A good agreement was found between the
two models, with Poynting fluxes of similar amplitude. In the 3D numerical
model, the non-linear interaction between the orbiting planet and the
star and its wind leads to a slightly more elongated \hbindex{Alfv\'en wing}
cross-section along ${\bf v}_0$. The two opposite topologies of the
dipolar interaction were also shown to lead to radically different
properties of the magnetic interaction with this model, leading to at
least an order of magnitude changes in magnetic torque and \hbindex{Poynting
flux}. For both aspects, these simulations helped realized how much the
effective area of the interaction (or, if one prefers, the obstacle)
significantly change with the topology. In the closed magnetosphere case (anti-aligned
configuration), the magnetospheric size can be well approximated using
the pressure ratio between the magnetospheric magnetic pressure on the
planetary side, and the total (thermal plus magnetic plus
\textit{ram}) pressure on the \hbindex{stellar wind} side of the
interaction. Assuming a spherical magnetosphere composed of a dipole
field, this leads to the well know expression 
\begin{equation}
  \label{eq:pressure_bal_magnetosphere}
  R_{\rm obst} = R_P \Lambda_P^{1/6} =
  R_P\left(\frac{B_P^2}{8\pi\,P_t}\right)^{1/6}\, ,
\end{equation}
where the subscript $P$ denotes values at the planetary surface and
$P_t$ is the total pressure of the \hbindex{stellar wind} at the planetary
orbit. This simple estimate nevertheless fails to describe the area of
interaction in the aligned case (see Figure \ref{fig:dipolar_numerical}), as in this case the
magnetic field lines connecting the planet and the wind that are part
of the \hbindex{Alfv\'en wings} act as well as an obstacle. As a
result, the area of interaction is far larger in the aligned case. In the
extreme case where the travel time of alfv\'enic perturbations is short
compared to the orbital period (see Eq. \ref{eq:travel_back_forth}), the waves can propagate back and forth
between the planet location and the stellar surface and the effective
area of interaction is the full \hbindex{Alfv\'en wing} from the planet location
to the stellar surface \citep[see also][]{Fleck:2008bp}. In general, though,
the effective obstacle will be only composed of a subpart of the \hbindex{Alfv\'en
wings} (see Figure 5 in \citealt{Strugarek:2016ee}).

Relying on the good agreement between the numerical and analytical
models, a parameter study using a self-consistent \hbindex{stellar wind} model
was undertaken in \citet{Strugarek:2016ee}. By changing the orbital
radius and magnetic field strength of the planet, empirical scaling laws have been
derived from a large set of non-linear numerical simulations for the
\hbindex{Poynting flux} and magnetic torque associated with
\hbindex{SPMI}. They can be summarized as follows (see bottom panels
in Figure \ref{fig:dipolar_numerical} and
\citealt{Strugarek:2016ee})

\begin{eqnarray}
  \label{eq:scaling_law_torque}
  \mathcal{T} &\propto& \left( c_d P_t M_a^\beta
  \right) \cdot \left(R_P^2R_{\rm orb}
\right)\cdot\left(\Lambda_P^\alpha\right) \, ,
  \\
  \mathcal{P} &\propto& \left( c_d S_w M_a^\xi
  \right) \cdot \left(R_P^2
\right)\cdot\left(\Lambda_P^\chi\right) \, .
  \label{eq:scaling_law_Poy}
\end{eqnarray}
We recall here that the coupling coefficient is defined as
$c_d=(4\pi/c^2)\Sigma_Av_o$ and the wind \hbindex{Poynting flux} is
$S_w=v_oB_w^2/(4\pi)$. The interested reader may find more
detailed scaling laws in \citet{Strugarek:2016ee}, including their
dependancy on the resistive properties of the plasma and of the
modelled ionospheric layer. The exponents ($\alpha,\beta,\xi,\chi$) vary with the topology of
the interaction and are given in \citet{Strugarek:2016ee}. The various
terms in Equations
\ref{eq:scaling_law_torque}-\ref{eq:scaling_law_Poy} have been
rearranged in three blocks from left to right: terms that depend only
on the star and its wind; only on the planet properties; and on a
combination of both ($\Lambda_P$). As a result these scaling laws may help relate
observed anomalous activity on distant stars
\citep[\textit{e.g.}][]{Shkolnik:2008gw} with the magnetic
properties (strength, topology) of the close-in planet, provided the \hbindex{stellar wind}
and planetary radius can be inferred or constrained observationally.

\section{Conclusions}
\label{sec:conclusions}

The study of star-planet magnetic interactions is a young and promising
field of research. In this review we have focused our discussion on
the modelling efforts that have been undertaken by the community in the past
decades, motivated by the many intriguing phenomena observed in
exoplanetary systems. Rather than listing again the effects
initiated by magnetic interactions, we list here several routes of improvement
of the models that need to be followed in support of the future
observation missions dedicated to the characterization of exoplanets
and their magnetic properties:

\begin{itemize}
\item In the context of both the unipolar and dipolar interaction
  cases, more realistic models of the interior of planets and their
  magnetosphere are needed. For example, an accurate response of the
  magnetospheric system to the impacting \hbindex{stellar wind} is needed to
  assess aurorae and possible planetary emissions, to assess the
  steady state of interaction and the energy conversion the
  interaction is able to operate, and to constrain the properties of
  the hypothetical bow-shock at the nose of the interaction. These
  improvements can be 
  carried out as a first step \textit{e.g.} by considering a
  magnetosphere-ionosphere coupling model in the case of the dipolar
  interaction, and more realistic planetary interior models in the
  context of the unipolar interaction. Ultimately, models including
  kinetic effects (beyond the standard
  magneto-hydrodynamic framework) will be needed for some of these aspects.
\item For the sake of simplicity, models have so far generally considered
  circular planetary orbits. This is often justified as a reasonable
  approximation, as close-in planets are likely in a tidally-locked
  state. Tidal theory nevertheless predicts departures from this simple
  picture \citep{Mathis:2017un}. As a result, eccentric orbits should be
  more systematically included in star-planet magnetic interaction models.
\item With a growing sample of stars hosting close-in exoplanets for
  which spectro-polarimetric observations are available, it is 
  today possible to simulate realistic \hbindex{stellar winds} of particular
  star-planet systems. Because of the variable nature of the magnetic
  interaction along the planetary orbit, it is essential to use these
  observational constraints to model close-in
  systems in order to assess the robustness of
  the simplified models, and quantitatively test our understanding of
  magnetic stat-planet interactions.
\item Last but not least, a significant effort has to be made to self-consistently
  compare magnetic effects with other physical mechanisms at stake in
  star-planet interactions. These include, but are not limited to,
  tides (torque, heating), radiation (ionization, atmospheric escape), and
  particle acceleration in the planetary magnetosphere (magnetic
  reconnection, instabilities). 
\end{itemize}

We hope this review will arouse multiple interests on
this promising and multi-disciplinary subject of research, and will encourage
further efforts in developing models of \hbindex{SPMI}s in all their complexity
to provide critical insights for the future observations of (close-in)
exoplanetary systems.

\section{Cross-References}

\begin{itemize}
\item{Electromagnetic Coupling in Star-Planet Systems}
\item{Magnetic Fields in Planet Hosting Stars}
\item{Tides in Star-Planet systems}
\item{Stellar Coronal and Wind Models: Impact on Exoplanets}
\item{Star-Planet Interactions in the Radio Domain: Prospect for Their Detection}
\item{Planetary Evaporation Through Evolution}
\item{Signatures of Star-Planet Interactions}
\item{Rotation of Planet-Harbouring Stars}
\item{Dynamical Evolution of Planetary Systems}
\item{Planetary Habitability and Magnetic Fields}
\end{itemize}

\begin{acknowledgement}
A. Strugarek acknowledges enlightening discussions about star-planet
interactions with A.S. Brun, J. Bouvier, D. C\'ebron, A. Cumming, S. Matt,
V. R\'eville, and P. Zarka. This review was written while A. Strugarek
was partially supported by the Canada's Natural Sciences and Engineering
Research Council, the ANR Blanc 2011 Toupies, 
and the Centre National d'Etudes Spatiales.
\end{acknowledgement}

\end{document}